\documentclass{jfm}

\usepackage{graphicx}
\usepackage{newtxtext}
\usepackage{newtxmath}
\usepackage{ragged2e}
\usepackage{blindtext}
\usepackage{natbib}
\usepackage{xcolor}
\usepackage{soul}
\usepackage{hyperref}
\hypersetup{
colorlinks = true,
urlcolor   = blue,
citecolor  = blue,
}

\DeclareUnicodeCharacter{2009}{\,} 
\usepackage[format=plain]{caption}
\makeatletter
\def\pagelimitfooter{}
\makeatother

\title{Ultrasound-controlled stream splitting in a microfluidic coflow}
\author{
D. Ghosh\aff{1},
S. Z. Hoque\aff{1,2},
T. Sujith\aff{1},
N. S. Satpathi\aff{1},
L. Malik\aff{1,3}
\and
A. K. Sen\aff{1}
\corresp{\email{ashis@iitm.ac.in}}
}

\affiliation{
\aff{1} Micro Nano Bio Fluidics Unit, Department of Mechanical Engineering, Indian Institute of Technology Madras, Chennai-600036, India

\aff{2} Current Affiliation: Department of Mechanical Engineering, Indian Institute of Technology Guwahati, North Guwahati - 781039, India

\aff{3} Current Affiliation: Department of Mechanical Engineering, Indian Institute of Technology Gandhinagar, Gujarat - 382055, India
}

\begin{document}
\maketitle

\thispagestyle{empty}

\begin{abstract}

Precise control of multiphase microfluidic flows underpins applications ranging from chemical processing to biomedical diagnostics. We investigate the response of a liquid--liquid coflow in a rectangular microchannel to an externally applied standing acoustic field. Acoustic excitation destabilizes an otherwise stable interface, giving rise to a sequence of reversible interfacial regimes: waviness, splitting, relocation, and stream-droplet breakup. Remarkably, a distinct splitting regime emerges, where a continuous stream partially splits into droplets at tunable locations while retaining a thin residual stream. Unlike conventional droplet breakup, this regime avoids complete disruption of the main flow, enables droplet generation at high capillary numbers, and allows spatial control over droplet formation. Extending across a broad range of capillary numbers, we examine how variations in flow conditions and applied acoustic power influence these regimes. Combining experiments, numerical simulations, and theoretical scaling, we elucidate the mechanisms governing this droplet generation mode and the associated regime transitions. Systematic measurements show that droplet size and residual stream thickness are governed primarily by hydrodynamic parameters, whereas the acoustic field controls the onset and spatial location of the breakup. These results establish a simple avenue for stream splitting and drop generation on-demand in a microfluidic coflow, opening new possibilities for spatially programmable manipulation of multiphase flows.

\end{abstract}



\section{Introduction}
\label{sec:1}

Understanding interfacial perturbations and their subsequent evolution is crucial for deciphering numerous natural phenomena involving fluid--fluid interfaces \citep{Stone2010,AlHousseiny2012}. From breaking waves on the shore to ink flowing from a pen, the geometric scales, fluid inertia, and dissipative mechanisms collectively determine the dominant instability mechanism. In recent years, microfluidic flows and associated interfacial instabilities have gained significant attention due to their wide-ranging applications in chemistry, biology, and materials science \citep{Gharib2022, Anna2016}. Within microchannels, fluid behaviour is highly sensitive to small perturbations, which can evolve into periodic interfacial waves or droplets depending on the underlying instability mechanism and flow conditions \citep{PhysRevLett.100.014502, dOLCE2009,Moragues2023,Kaminski2016}. For instance, passive co-flow systems have been shown to develop periodic interfacial waves when inertia and viscosity contrast become sufficiently large \citep{Hu2018,Yih_1967}. Precise control and manipulation of such interfacial structures, particularly droplets, is essential for a wide range of microfluidic applications including emulsification, encapsulation, chemical synthesis, and biomedical assays \citep{Shang2017}.
Conventionally, droplet generation in microchannels is achieved using T-junctions \citep{Garstecki2006,Xu2006} or flow-focusing geometries \citep{Anna2003,Anna2003a}, where droplet formation is typically governed by an absolute instability \citep{Guillot2007}. In these configurations, the dispersed phase is squeezed by the continuous phase at a confined junction, leading to droplet breakup controlled primarily by geometric confinement and the flow-rate ratio. Although widely adopted, such junction-based configurations rely on narrow constrictions or nozzles that complicate fabrication, increase the risk of clogging, and restrict the range of accessible droplet sizes \citep{Garstecki2006,Dewandre2020,Malik2024}. Recent refinements of the classical squeezing regime have further revealed intrinsic limitations of junction-based droplet generation. In cross-junction microchannels operating at very low capillary numbers $(Ca)$, deviations from the conventional squeezing law have been attributed to the emergence of a leaking regime, wherein the continuous phase partially bypasses the forming droplet through corner gutters, rendering both the filling and necking stages sensitive to capillary number \citep{Kurniawan2023}. Despite such mechanistic refinements, droplet formation in these systems remains fundamentally tied to geometric confinement at the junction, and stable parallel flow persist at sufficiently high capillary numbers. An alternative approach involves the generation of droplets driven by convective instability \citep{Anna2003,Nunes2013,Cramer2004}, where a jetting mechanism enables the formation of droplets without relying strictly on geometric squeezing. However, achieving stable and continuous droplet generation while maintaining control over droplet size and morphology remains challenging due to the inherently unstable nature of convective jetting \citep{Nan2024}. 

Passive microfluidic methods have also been explored to exploit interfacial instabilities between co-flowing immiscible fluids in microchannels  \citep{Hu2018,Hazra2024,Cubaud2008}. Nevertheless, these approaches typically require high viscosity contrasts or operation at very low capillary numbers $(Ca \ll 1)$, introducing additional design constraints. Moreover, at moderate to high capillary numbers $(Ca \gtrsim 1)$, interfacial disturbances are rapidly convected downstream before they can amplify, resulting in stable parallel coflow without droplet breakup \citep{Guillot2008}. Consequently, generation of droplets in microchannels at such capillary numbers $(Ca \gtrsim 1)$, where the base coflow is hydrodynamically stable, remains largely unexplored.

\begin{figure}
\centering
\includegraphics[clip,width=1\textwidth]{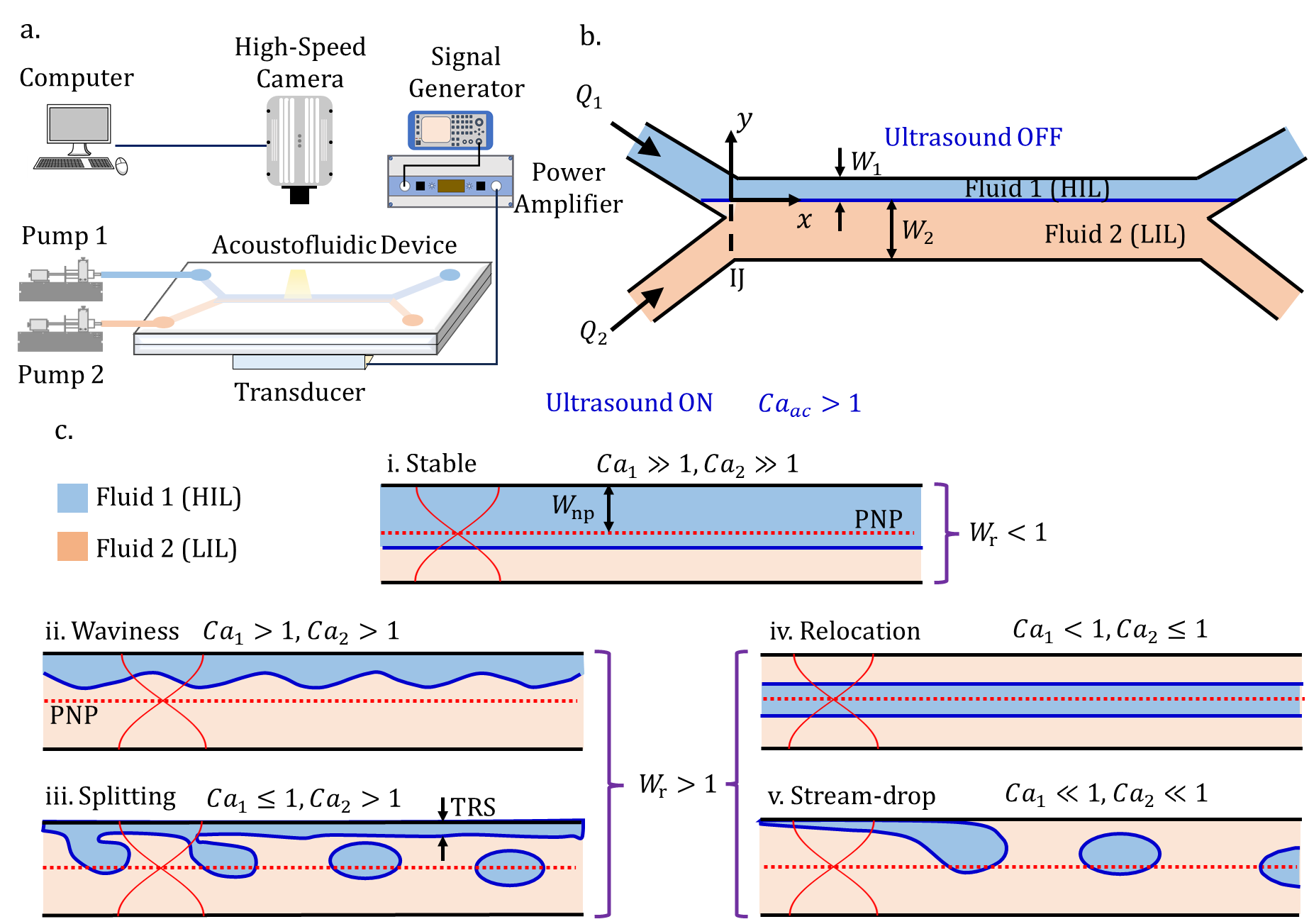}
\caption
{
\justifying{ (a) Schematic of the experimental setup (top left). The system consists of a silicon--glass microchannel with two inlets and two outlets forming a straight channel to establish a coflow configuration. A planar piezoelectric transducer is attached beneath the device to generate an ultrasonic standing-wave field. The transducer is driven by a signal generator connected to a power amplifier. Syringe pumps control the infusion of the working fluids, and a high-speed camera records the interfacial dynamics. Subscripts 1 and   2 in the flow rate ($Q$) and stream width ($W$) denote the high-impedance liquid (HIL) and low-impedance liquid (LIL), respectively. (b) Schematic of a stable coflow system when the ultrasound field is OFF. The inlet junction (IJ) marks $x=0$ from which the downstream axial distance ($x$) is measured. (c) When the ultrasound field is ON, distinct flow regimes emerge, categorized based on the capillary number ($Ca$) and the variable $W_r = W_{np}/W_1$, where $W_{np}$ is the distance of the acoustic nodal plane from the HIL-side wall and $W_1$ is the HIL stream width. The standing half-wave configuration is indicated by a red continuous line and the nodal line by a red dash--dot line. The observed regimes include (i) stable coflow, (ii) interfacial waviness, characterized by periodic interfacial deformation, (iii) stream splitting, where the HIL splits into droplets and a thin residual stream (TRS), (iv) stream relocation, where the HIL relocates to the pressure nodal plane at the channel centre, and (v) stream-to-drop breakup, corresponding to complete breakup of the HIL stream into droplets. } 
}
\label{fig1}
\end{figure}

Due to the limitations of passive strategies for interfacial manipulation, recent studies have explored external field-based approaches to actively control droplet generation \citep{Shang2017}. The review of droplet microfluidics highlights a wide range of actuation strategies, including electric, optical, thermal, and acoustic forcing \citep{Nan2024}. Electric-field-induced droplet ejection, piezo-driven modulation, and pressure-controlled on-demand generators allow precise temporal control of droplet formation by applying localized forcing at junctions or nozzles \citep{Lin2012}. Similarly, laser-induced cavitation has been employed to trigger droplet pinch-off via transient pressure pulses generated by bubble dynamics \citep{Park2011}. Optocapillary forcing has also been used to locally pinch confined threads and induce a transition from convective to absolute instability within microchannels \citep{SaintVincent2016}. Although these active approaches provide valuable temporal control over droplet emission, they generally rely on localized actuation near geometric constrictions or operate in regimes where the base flow is already susceptible to breakup $(Ca \ll 1)$. Evidently, generation of droplets in microchannels at moderate to high capillary numbers $(Ca \gtrsim 1)$ with stable base coflow remains a challenge.
Among active techniques, ultrasound-based methods have attracted considerable attention owing to their non-invasive character, biocompatibility, and label-free operation \citep{Li2018}. The acoustic radiation force generated by acoustic impedance contrast between immiscible fluids can modify interfacial stresses and relocate co-flowing streams \citep{Hemachandran2019,Deshmukh2014,Nath2021,Zhu2017,Karthick2018}. Acoustic vibrations in the kilohertz (kHz) range have been applied to liquid jets with ultra-low interfacial tension, producing interfacial folding and droplet formation \citep{Mak2017}. However, these approaches typically produce droplets of fixed size, lack control over the breakup location, or are not readily adaptable to confined microchannels. Recently, exposure of an immiscible coflow system to bulk acoustic waves enabled reversible transition between continuous and droplet modes, at a fixed $Ca$ \citep{Hemachandran2021}. Nevertheless, droplet generation at high $Ca$ remained unattainable because ultrasound actuation led to relocation of the entire fluid stream rather than destabilizing the interface. Beyond droplet generation, the formation of thin liquid layers inside microfluidic channels is of considerable interest for applications such as coating, lubrication, surface patterning, and fiber fabrication. Microfluidic strategies for producing thin liquid films typically rely on hydrodynamic focusing or carefully tuned flow-rate ratios \citep{Andrieux2021,Ouyang2023}. These approaches often require precise flow control or specialized channel geometries, motivating the development of alternative mechanisms for controllable thin-layer generation in microfluidic systems.
\\
\\
In the present work, we study the acoustic forcing of an otherwise stable immiscible coflow system, which gives rise to three distinct interfacial phenomena. First, we report a previously unexplored splitting regime in which a continuous liquid stream divides into a train of droplets and a thin residual stream. Remarkably, this regime enables droplet generation at moderate to high capillary numbers $(Ca \gtrsim 1)$, where parallel coflows are found to be hydrodynamically stable. The breakup location and droplet size can be tuned through the acoustic field strength, enabling spatially controlled droplet generation. Second, the splitting mechanism inherently produces a thin residual stream along the channel wall, providing a new avenue for the generation of thin liquid films within microfluidic devices. Third, we identify a waviness regime in which the interface develops spatially periodic oscillations solely due to acoustic forcing. Unlike previously reported interfacial waviness driven by inertia or high viscosity contrasts \citep{Hu2018}, the interfacial waves observed here arises even at low Reynolds number and low viscosity contrast, suggesting that the interface destabilization is attributed to acoustic stresses. By systematically varying capillary number and acoustic forcing conditions, we observe transitions between several distinct regimes: stable coflow, inetracial waviness, stream splitting, stream relocation, and stream-to-drop breakup (see Figure \ref{fig1}). Through a combination of experiments, numerical simulations, and theoretical scaling analysis, we elucidate the underlying force balance between acoustic radiation force, viscous force, and interfacial tension that governs these regime transitions and the emergence of acoustically induced interfacial instabilities.

\section{Experimental}\label{sec:2}

\subsection{Device fabrication and experimental setup}\label{sec:2.1}

Experiments are performed in a silicon–glass device comprising a straight rectangular microchannel of width $W = 370~\mu\text{m}$ and height $h = 100~\mu\text{m}$, with two inlets and two outlets (see Figure~\ref{fig1}(b)). The microchannel device is made up of a 4-inch, single-side polished $\langle 100 \rangle$ silicon wafer of thickness $500~\mu\text{m}$. The channel pattern is defined using ultraviolet (UV) photolithography, followed by deep reactive ion etching (DRIE) to a depth of $100~\mu\text{m}$. After photoresist removal and cleaning, the patterned silicon wafer is anodically bonded to a $500~\mu\text{m}$-thick borosilicate glass wafer at $450^{\circ}\text{C}$ under an applied potential of 1000~V, forming a sealed microchannel with optical access through the glass layer. A schematic of the experimental setup is presented in Figure~\ref{fig1}(a). Fluid delivery is achieved using a high-precision, pulsation-free syringe pump (neMESYS, Cetoni GmbH, Germany) connected to the device via PTFE tubing. The device is mounted on an upright optical microscope (Axiovert, Zeiss, Germany) equipped with long-working-distance objectives (10X and 20X) for visualization through the transparent glass substrate. A high-speed camera (FASTCAM SA5, Photron, Japan) operating at 1000~fps (frames per second) captures the dynamics of the interface. 

Acoustic excitation is applied using a planar piezoceramic transducer (PZT, Sparkler Piezoceramics, India) bonded beneath the silicon substrate using a thin layer of epoxy to enhance acoustic coupling and minimize energy losses. The PZT is driven by a sinusoidal signal generated using a radio-frequency (RF) signal generator (SMC100A, Rohde \& Schwarz), amplified using a broadband power amplifier (75A100A, Amplifier Research). The acoustic power ($P$) can be controlled by adjusting the gain of the signal generator and power amplifier. We measured peak-to-peak voltage ($V_{pp}$) using an Oscilloscope, which is a measure of the input acoustic power. The acoustic wave generated by the PZT transducer propagates through the silicon substrate and into the fluid-filled microchannel, generating a symmetric acoustic pressure field across the channel. The driving frequency ($f$) is varied between 1.90--2.25~MHz to achieve resonance conditions, and the peak-to-peak voltage is varied in the range 0--10~V depending on the fluid combination and the desired flow regime.

\subsection{Materials and fluid properties}\label{sec:2.2}

The liquids used in this study were obtained from Sigma-Aldrich (Bangalore, India). Before experiments, each liquid was filtered through a $0.45~\mu\text{m}$ nylon syringe filter to remove particulates and degassed to remove dissolved gases. The speed of sound in each liquid was measured using an ultrasonic interferometer (Mittal Enterprises, India). The static contact angles of the different working liquids on clean polished Si (100) wafers were measured using a goniometer (Krüss DSA25, Germany) following standard cleaning procedures and nitrogen drying. The interfacial tension (IFT) between relevant fluid pairs forming coflows was determined using the pendant-drop method. The measured properties of the different working fluids are summarised in Table~\ref{tab:tab1}, and interfacial tensions for relevant fluid combinations are presented in Table~\ref{tab:tab2}. All reported measurements represent averages over multiple trials and lie within the specified uncertainty.

\begin{table}
\centering
\caption{Properties of the different fluids used in the experiments.}
\begin{tabular}{lcccccc}
\hline
Fluid & Density & Speed of sound & Impedance & Viscosity & Contact angle & Surface tension \\
 & (kg/m$^3$) & (m/s) & ($\times 10^6$ kg/m$^2$s) & (mPa$\cdot$s) & ($^\circ$) & (mN/m) \\
\hline
Olive oil & 917 & 1452 & 1.331 & 62.3 & $23\pm 2$ & 32 \\
Mineral oil & 860 & 1412 & 1.214 & 860 & $16\pm 1$ & 30 \\
Silicon oil 100 & 975 & 1032 & 1.006 & 105.6 & $3.2\pm 1$ & 21 \\
Silicon oil 350 & 946 & 1010 & 0.955 & 355.4 & $2.5\pm 1$ & 20 \\
\end{tabular}
\label{tab:tab1}
\end{table}


\begin{table}
\centering
\caption{Interfacial tension between relevant fluid combinations.}
\begin{tabular}{lll}
\hline
Liquid 1 (LIL) & Liquid 2 (HIL) & Interfacial tension (mN/m) \\
\hline
Silicon 100 & Olive oil & $0.913 \pm 0.10$ \\
Silicon 350 & Olive oil & $1.174 \pm 0.10$ \\
Silicon 100 & Mineral oil & $0.540 \pm 0.20$ \\
Silicon 350 & Mineral oil & $0.462 \pm 0.15$ \\
\\
\end{tabular}
\label{tab:tab2}
\end{table}


\section{Numerical models}\label{sec:3}

To elucidate the underlying physics of the ultrasound-actuated coflow system, numerical simulations were performed using COMSOL \textit{Multiphysics}, which provides a finite-element framework for solving the coupled acoustic, structural and fluid dynamical equations \citep{Bruus2012-gd,Hoque2023,Malik2025-bl,Sujith2024}. The numerical modeling framework is organized as follows. First, the governing equations for the acoustic response are formulated and reduced to a Helmholtz equation for the first-order acoustic pressure field. This equation is then solved in a reduced three-dimensional coflow domain to determine the resonance modes and the corresponding position of the pressure nodal plane. However, the reduced-order model does not capture the axial variation of the acoustic field arising from the actual electromechanical coupling between the piezoelectric transducer and the microfluidic device. Therefore, fully coupled (electromechanical) multiphysics simulations are performed for selected cases at the resonance modes identified from the reduced-order analysis. Finally, since the acoustic field shows negligible variation along the channel height, transient two-phase simulations are carried out in a reduced two-dimensional domain to resolve the interfacial dynamics and reproduce the experimentally observed deformation regimes. The details of these models are given below.

\subsection{Governing equations for the acoustic field}\label{sec:3.1}

The acoustic field in a compressible Newtonian fluid is governed by the conservation of mass and momentum, together with an equation of state. The pressure $p$, density $\rho$ and velocity $\mathbf{u}$ satisfy
\begin{equation}
p = p(\rho),
\end{equation}
\begin{equation}\label{eqn~3.2}
\frac{\partial \rho}{\partial t} = - \nabla \cdot (\rho \mathbf{u}),
\end{equation}
\begin{equation}\label{eqn~3.3}
\rho \frac{\partial \mathbf{u}}{\partial t} + \rho (\mathbf{u} \cdot \nabla)\mathbf{u}
= -\nabla p + \mu \nabla^2 \mathbf{u} + \beta_r \mu \nabla (\nabla \cdot \mathbf{u}) + \mathbf{F}_b,
\end{equation}
where $\mu$ is the dynamic viscosity, $\beta_r$ is the ratio of bulk to dynamic viscosity, and $\mathbf{F}_b$ denotes the body force acting on the fluid.

To describe acoustic actuation, small perturbations about a quiescent state are introduced as $\wp = \wp_0 + \wp_1 + \wp_2 \cdots$, where $\wp$ represents any variable, such as pressure, velocity, or density. Here, $\wp_0$ corresponds to the equilibrium state, and $\wp_1$ denotes the first-order acoustic perturbation field. These subscripts $1$ and $2$ should not be confused with the subscripts used later to distinguish the two liquids.

For adiabatic acoustic propagation, the first-order pressure perturbation is related to the density perturbation through $p_1 = c_0^2 \rho_1$, where $c_0$ is the isentropic speed of sound in the fluid. Linearising eqns.~\ref{eqn~3.2} and~\ref{eqn~3.3}, with second-order terms neglected, and assuming time-harmonic variations of the perturbation quantities such that any first-order variable $\wp_1(\mathbf{r},t)$ can be expressed as $\wp_1(\mathbf{r},t) = \hat{\wp}_1(\mathbf{r}) e^{-i\omega t}$, where $\omega$ is the angular frequency of excitation, yields the frequency-domain formulation.

Substituting these time-harmonic forms into the linearised governing equations yields the governing equation for the first-order acoustic pressure field,
\begin{equation}\label{Eqn-3.4}
\nabla^2 p_1 = -k_0^2 p_1,
\end{equation}
where $k_0=\omega/c_0$ is the acoustic wavenumber, with $\omega$ denoting the angular frequency. Equation~\eqref{Eqn-3.4} forms the basis for determining the acoustic response of the coflow system.

\subsection{Determination of the pressure nodal plane}\label{sec:3.2}

To determine the position of the pressure nodal plane in the coflow configuration, a reduced three-dimensional domain representing a section of the microchannel is considered (see Appendix~\ref{appA}). This reduced geometry retains the essential acoustic features while significantly lowering the computational cost. The channel has a total width $W=W_1+W_2=370~\mu\mathrm{m}$, where $W_1$ and $W_2$ correspond to the widths of the high-impedance liquid (HIL) and low-impedance liquid (LIL), respectively, obtained from the experimental flow conditions. The channel height is $h=100~\mu\mathrm{m}$ and the computational length is $L=5~{mm}$. To obtain the nodal position, we follow the numerical approach previously used for evaluating first-order acoustic fields in microfluidic coflow systems \citep{Hoque2023}. The acoustic pressure distribution inside the three-dimensional coflow domain is computed by solving the Helmholtz equation~\eqref{Eqn-3.4} using the Pressure Acoustics module in COMSOL \textit{Multiphysics}.

In the coflow configuration, a plane acoustic wave propagating through one fluid interacts with the liquid--liquid interface and undergoes partial reflection and transmission due to the acoustic impedance mismatch between the two phases. At the liquid--liquid interface, the acoustic pressure remains continuous across the interface to ensure continuity of the normal stress. Under these conditions, the background acoustic pressure field, composed of incident and reflected waves, can be expressed as \citep{Baasch2020,Hoque2023}
\begin{equation}\label{Eqn-3.2}
p_1 = p_{\mathrm{in}}\left[\overline{\Delta Z}\cos(k_0y+\omega t) + \cos(\omega t-k_0y)\right],
\end{equation}
where $p_{\mathrm{in}}$ represents the amplitude of the incoming background pressure wave. In typical acoustofluidic devices, under resonance conditions, the acoustic pressure amplitude is commonly of the order of $p_{\mathrm{in}} \sim 1~\mathrm{MPa}$ \citep{Hoque2023}. The parameter $\overline{\Delta Z}=(Z_1-Z_2)/(Z_1+Z_2)$ represents the reflection coefficient arising from the acoustic impedance mismatch, where $Z_1$ and $Z_2$ denote the acoustic impedances of the high-impedance and low-impedance liquids, respectively.

The acoustic boundary conditions at the channel walls depend on the acoustic impedance contrast between the fluid and the solid substrate. In the present system, the acoustic impedances of silicon and borosilicate glass are significantly larger than those of the liquids. Consequently, the channel walls can be approximated as acoustically rigid boundaries, leading to the hard-wall boundary condition
\begin{equation}\label{Eqn-3.3}
\mathbf{n}\cdot\nabla p_1 = 0 ,
\end{equation}
where $\mathbf{n}$ denotes the outward normal vector to the channel wall.

The Helmholtz equation (eqn.~\eqref{Eqn-3.4}), together with the interface continuity condition and the hard-wall boundary condition~\eqref{Eqn-3.3}, is solved in an eigenfrequency framework to obtain the acoustic resonance modes of the coflow system. From the resulting pressure field, the position of the pressure nodal plane is extracted as a function of the fluid configuration. Although this reduced three-dimensional eigenfrequency analysis provides the resonance modes and the corresponding nodal-plane location, it does not capture the axial variation of the acoustic pressure field along the microchannel. This axial variation originates from the actual electromechanical coupling between the piezoelectric transducer and the silicon--glass microfluidic device containing the microchannel. Therefore, fully coupled electromechanical simulations are performed for selected cases near the resonance conditions identified from the reduced-order model, as described below.

\subsection{Fully-coupled electromechanical simulation}\label{sec:3.3}

To determine the spatial variation of the acoustic pressure field within the complete acoustofluidic device, a fully-coupled numerical model was developed in COMSOL \textit{Multiphysics}. The model incorporates the coupled electromechanical interaction between the piezoelectric transducer, the silicon--glass device, and the fluid domain inside the microchannel (see Appendix~\ref{appA}). The model combines three physics interfaces in COMSOL \textit{Multiphysics}: the Pressure Acoustics module for the fluid domain, the Solid Mechanics module for the structural components of the device, and the Electrostatics module for the actuation of the piezoelectric transducer \citep{Hoque2022}. The acoustic--structure boundary interface couples the pressure acoustics and solid mechanics domains, while the piezoelectric multiphysics interface couples the solid mechanics and electrostatics modules within the PZT layer. The electromechanical simulations are performed for selected cases near the resonance conditions identified from the reduced three-dimensional model in \S~\ref{sec:3.2}. This approach significantly reduces computational cost because the resonance modes are first identified using the reduced-order eigenfrequency analysis, rather than by attempting to compute the entire device response directly from the fully-coupled electromechanical model. The full-chip simulation is then used to predict the axial variation of the pressure nodal plane.

Since the vibration amplitudes of the device components are small compared to the device dimensions, the problem is solved in the frequency domain using time-harmonic fields, consistent with the formulation introduced in \S~\ref{sec:3.1}. The elastic deformation of the structural domains is governed by the frequency-domain form of the Cauchy equation of motion
\begin{equation}
\nabla \cdot \boldsymbol{\sigma_s} + \rho_s \omega^2 \mathbf{v}_s = 0 ,
\end{equation}
where $\boldsymbol{\sigma_s}$ is the stress tensor, $\rho_s$ is the density of the solid material, and $\mathbf{v}_s$ is the displacement field. The strain tensor $\boldsymbol{E}$ is related to the displacement field using
\begin{equation}
\boldsymbol{E}=\frac{1}{2}\left[\nabla \mathbf{v}_s + (\nabla \mathbf{v}_s)^T \right].
\end{equation}
The stress--strain relation for linear elastic materials is expressed as
\begin{equation}
\boldsymbol{\sigma_s} = \mathbf{C} : \boldsymbol{E},
\end{equation}
where $\mathbf{C}$ is the elasticity tensor. In the present model, silicon is treated as an anisotropic elastic material, whereas the borosilicate glass layer is modeled as isotropic. The corresponding elastic constants and material properties are taken from reported literature values \citep{Hoque2022}.

The electromechanical behavior of the PZT transducer is described using the coupled piezoelectric relations
\begin{equation}
\nabla \cdot \mathbf{D} = 0 ,
\end{equation}
\begin{equation}
\boldsymbol{\sigma_s} = \mathbf{C}^E : \boldsymbol{E} + \mathbf{d_p}^T \cdot \nabla V_E ,
\end{equation}
\begin{equation}
\mathbf{D} = \mathbf{d_p} : \boldsymbol{E} - \varepsilon_0 \boldsymbol{\varepsilon}_r^S \nabla V_E ,
\end{equation}
where $V_E$ is the electric potential, $\mathbf{D}$ is the electric displacement field, $\mathbf{d_p}$ is the piezoelectric coupling tensor, $\mathbf{C}^E$ is the elastic stiffness tensor at constant electric field, $\varepsilon_0$ is the permittivity of free space, and $\boldsymbol{\varepsilon}_r^S$ is the relative permittivity tensor at constant strain. The material parameters of the PZT transducer are taken from standard literature.

In the fluid domain, the acoustic pressure field is obtained by solving the first-order Helmholtz equation (eqn.~\eqref{Eqn-3.4}). The resulting multiphysics model captures the electromechanical excitation of the device and the resulting acoustic pressure distribution within the microchannel geometry containing the silicone oil--olive oil coflow. The simulated pressure field is subsequently used to determine the axial variation of the pressure nodal plane within the microchannel.

\subsection{Dynamic simulation of the interface evolution}\label{sec:3.4}

The acoustic-field analysis presented in \S~\ref{sec:3.2} shows negligible variation of the acoustic pressure along the channel height ($z$-direction); see supplementary material S1. Therefore, transient simulations are carried out in a two-dimensional domain ($xy$-plane) (see Appendix~\ref{appA}), which retains the essential physics of the interface evolution while significantly reducing the computational cost.

To reproduce the experimentally observed interfacial deformation patterns, including the splitting and waviness regimes, time-dependent two-phase simulations \citep{Hoque2024} were performed using the Laminar Flow and Phase-Field modules in COMSOL \textit{Multiphysics}. The simulations capture the coupled hydrodynamic and interfacial response of the two immiscible liquids under acoustic forcing.

The two-phase flow is governed by the conservation of mass and momentum, as described in \S~3.1. The flow is assumed to be incompressible, $\nabla \cdot \mathbf{u} = 0$. The acoustic radiation force is incorporated as a volumetric body force through the $\mathbf{F}_v$ term in equation~\ref{Eqn-3.4}. To correctly represent the spatial variation of fluid properties across the interface, the density and viscosity are defined in terms of the local volume fraction of the HIL as $\rho=\rho_2+(\rho_1-\rho_2)V_{f1}$ and $\mu=\mu_2+(\mu_1-\mu_2)V_{f1}$, where $V_{f1}=\min(\max((1-\phi)/2,0),1)$ and the local volume fraction of the LIL is $V_{f2}=1-V_{f1}$, $\phi$ is the phase-field variable.

The interface separating two immiscible liquids is captured using the Phase-Field Method. In this formulation, the interface is represented by a phase-field variable $\phi$, which takes nearly constant values in the bulk phases and varies smoothly across a finite-thickness transition layer corresponding to the diffused interface. This approach eliminates the need for explicit interface reconstruction and naturally accommodates large interfacial deformations and topological changes such as breakup. In contrast to sharp-interface approaches, which often require additional slip-length models to regularise the moving contact line, the phase-field formulation captures the contact-line dynamics through the diffusion of free energy. Consequently, the phase-field method has been widely adopted in the literature for simulating immiscible two-phase flows in microfluidic systems \citep{Santraaaaaa2020}.

The evolution of the phase-field variable $\phi$ is governed by the convective Cahn--Hilliard equation
\begin{equation}\label{Eqn-3.6}
\frac{\partial \phi}{\partial t}+\mathbf{u}\cdot\nabla\phi
=\nabla\cdot\left(\frac{M\nu}{\varepsilon^2}\nabla\xi\right),
\end{equation}
where $M$ is the mobility parameter controlling the diffusion time scale of the interface, $\nu$ is the mixing energy density, and $\varepsilon$ is the capillary width associated with the thickness of the diffused interface. The free energy function is defined as $\xi=-\nabla\cdot(\varepsilon^2\nabla\phi)+(\phi^3-\phi)$.

The interfacial tension force arises from the phase-field free-energy function through the chemical potential $G=\nu\xi/\varepsilon^2$, and is incorporated as a body-force contribution $\mathbf{F}_I$ in the momentum equation via the term $\mathbf{F}_v$. The acoustic radiation force $\mathbf{F}_{ac}$ generated by the standing acoustic field is also included in $\mathbf{F}_v$, such that the total body force accounts for both interfacial and acoustic effects ($\mathbf{F}_v=\mathbf{F}_I+ \mathbf{F}_{ac}$).

The unit normal to the interface is obtained from the gradient of the phase-field variable $\phi$, ensuring that the force formulation remains valid throughout the domain and that curvature effects are naturally captured. The acoustic radiation force acting on the interface is expressed as
\begin{equation}\label{Eqn-3.7}
\mathbf{F}_{ac} = -\langle E_{ac} \rangle A_I
\left[ e^{i2k_2y} + \overline{\Delta Z}^2 e^{-i2k_2y}
- \left(\frac{\rho_1}{\rho_2}\right)(1-\overline{\Delta Z})^2 e^{i2k_1y} \right]\mathbf{n},
\end{equation}
where $\langle E_{ac}\rangle$ is the time-averaged acoustic energy density, $A_I$ denotes the interface area, and $k_1$ and $k_2$ denote the wavenumbers in the HIL and LIL, respectively.

To solve the phase-field transport equation, appropriate boundary conditions are imposed at the channel walls. The wetting behaviour of the liquids is enforced through the contact angle $\theta_c$, implemented using the phase-field boundary condition $\mathbf{n}\cdot\varepsilon^{2}\nabla\phi=\varepsilon^{2}\cos(\theta_c)|\nabla\phi|$. The contact angle, $\theta_c$, is defined as the angle between the liquid--liquid interface and the microchannel wall on the HIL side. The values of $\theta_c$ are taken from previously reported measurements for silicone oil--olive oil systems \citep{Hoque2024}. In addition, a zero-flux condition for free energy is applied as $\mathbf{n}\cdot (M\nu/\varepsilon^2)\nabla\xi=0$. No-slip boundary conditions are imposed at the channel walls, fully developed velocity profiles corresponding to the prescribed flow rates are applied at the inlets, and a constant pressure condition is prescribed at the outlet.

Acoustic streaming effects are neglected in the present simulations, since their magnitude is significantly smaller than the acoustic radiation force for liquid pairs with strong acoustic impedance contrast \citep{Hemachandran2021}. The resulting model captures the coupled effects of hydrodynamics, interfacial tension, and acoustic forcing responsible for the transition between the splitting and waviness regimes.

\section{Results and discussion}\label{sec:4}

We consider different coflow systems in a rectangular microchannel, each comprising a pair of immiscible high-impedance liquid (HIL) and a low-impedance liquid (LIL). We present results for systems where olive oil and mineral oil are used as the HIL (Fluid 1), while silicone oil 100 and 350 are used as the LIL (Fluid 2). The measured properties of these liquid combinations are presented in Tables~\ref{tab:tab1} and \ref{tab:tab2}. The interfacial dynamics are governed by the combined influence of flow rates, fluid viscosities, and interfacial tension between the liquids. To account for these effects in a unified manner, the results are presented in terms of the capillary number, defined for each phase as $Ca_1 = \mu_1 U_1/\gamma_{12}$ and $Ca_2 = \mu_2 U_2/\gamma_{12}$, where $\mu$, $U$, and $\gamma_{12}$ denote the dynamic viscosity, cross-sectionally averaged velocity, and interfacial tension, respectively. Subscripts 1 and 2 correspond to the HIL and LIL phases. In addition, the effect of acoustic forcing is characterized through the acoustocapillary number, $Ca_{ac}$, which is defined as the ratio of acoustic radiation force to interfacial tension force \citep{Hemachandran2021}, and is expressed as $Ca_{ac}=E_{ac}\overline{\Delta Z}W/\gamma_{12}$, where $E_{ac}$ energy density, $\overline{\Delta Z}$ normalized acoustic impedance contrast, $W$ channel width and $\gamma_{12}$ is interfacial tension. The flow dynamics are investigated by systematically varying the flow rates of the liquid pairs and the applied acoustic power, and the resulting interfacial responses are analysed in terms of the above dimensionless parameters.

\subsection{Stable base coflow in the absence of acoustic actuation}\label{sec:4.1}

We first establish a stable coflow of a pair of immiscible liquids inside a microchannel, in the absence of acoustic actuation. Experiments were performed over a wide range of flow rates (1--200~$\mu{L}min^{-1}$) of a pair of phases in order to identify the operating conditions under which a stable co-flow can be established prior to the application of ultrasound. The high-impedance liquid (HIL) and low-impedance liquid (LIL) occupy widths $W_1$ and $W_2$, respectively, and are separated by a flat interface with interfacial tension $\gamma_{12}$ (see Figure~\ref{fig1}). We observe that stable parallel coflow is not obtained for all combinations of $Ca_1$ and $Ca_2$. In particular, when the capillary number of one fluid is much smaller than that of the other, viscous shear stresses between the two streams become strongly imbalanced. As a result, the interface becomes unstable even without acoustic forcing, preventing the establishment of a steady coflow. We refer to these conditions as the no-coflow regime as shown in the regime map corresponding to the operating range in the absence of an acoustic field in Appendix~\ref{appB}. Consequently, our experiments are restricted to operating conditions where a pair of streams form an inherently stable coflow prior to acoustic actuation. For such stable co-flow conditions, the equilibrium position of the interface is determined by the stratified flow established in the channel. Assuming a fully developed, incompressible, and laminar flow in both phases, the interface location can be obtained from the analytical solution of the two-fluid flow field, which relates the stream widths to the flow-rate and viscosity ratios \citep{Hazra2024}. Following this formulation, the normalized width of the HIL stream can be approximated as
$W_1/W = 1 - (1 + 0.3 (Q_r \mu_r)^{1/3})^{-2}$,
where $Q_r = Q_1/Q_2$ is the flow-rate ratio and $\mu_r = \mu_1/\mu_2$ is the viscosity ratio, with $Q_1$ and $Q_2$ denoting the volumetric flow rates of the HIL and LIL streams, respectively. This relation provides an estimate of the equilibrium interface position for the hydrodynamically stable coflow prior to the application of ultrasound. The theoretical model for the width ratio and velocity profiles in case of the primary-flow solution is presented in Appendix~\ref{appC} (see supplementary material S2).

\begin{figure}
\centering
\includegraphics[clip,width=1\textwidth]{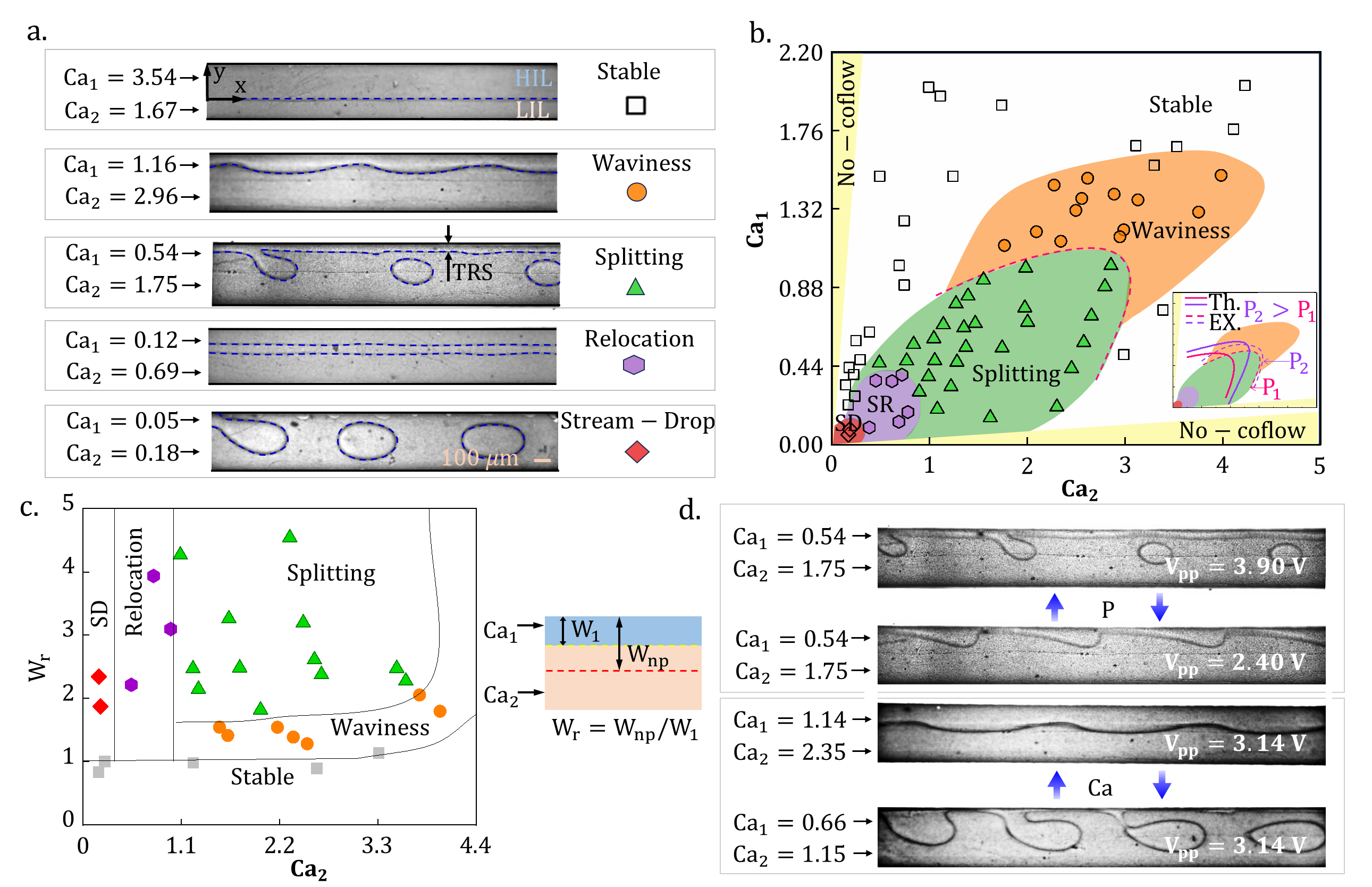}
\caption
{
\justifying{(a) Experimental images illustrating the different flow regimes: stable coflow, interfacial waviness, stream splitting, stream relocation (SR), and stream-drop (SD) breakup. (b) Regime map depicting the distinct flow patterns when an inherently stable coflow system is exposed to an ultrasound field, in terms of the Ca of the pair of liquids at fixed acoustic input power, $P_1$=125 mW. The symbols are shown in (a). Inherently stable coflow is not established in the yellow-shaded region. The inset shows the shift in the boundary between the splitting and waviness regimes at two different power levels ($P_1$=125 mW and $P_2$=220 mW). Continuous and dashed lines represent theoretical predictions and experimental observations, respectively. (c) Regime map  showing the different regimes in terms of the position of the pressure nodal plane, obtained from simulations, $W_r=W_{np}/W_1$ and $Ca_2$, here $W_{np}$ is the distance from the HIL-side wall to the pressure node and $W_1$ is the width of the HIL stream. (d) Experimental images depicting the reversible transition between the waviness and splitting regimes by controlling acoustic power (or $V_{pp}$)) at a fixed $Ca$ of the liquids and controlling $Ca$ at fixed $V_{pp}$. Experimental images correspond to the Silicon-Olive oil coflow system.}}
\label{fig2}
\end{figure}

\subsection{Onset of interfacial deformation upon ultrasound actuation and Regime transition}\label{sec:4.2}
Upon exposure of a stable coflow system to ultrasound actuation, the liquid--liquid interface undergoes deformation due to acoustic radiation forces generated by the scattering of the incident acoustic field at the interface. Experimentally, we observe that the initially stable interface exhibits a range of distinct responses depending on the flow conditions and the acoustic forcing, including stable co-flow, interfacial waviness, stream splitting, stream relocation, and stream-droplet breakup. These interfacial behaviours are governed by the combined influence of the capillary numbers of the liquid pair ($Ca_1$, $Ca_2$), the acoustocapillary number $Ca_{ac}$, and the relative position of the acoustic pressure node with respect to the liquid--liquid interface.

Our experiments show that a necessary condition for interfacial deformation is $Ca_{ac}\gtrsim1$, indicating that the acoustic radiation force must overcome the restoring interfacial tension force. Once this condition is satisfied, the nature of the instability is governed primarily by the capillary numbers of both liquids. Representative schematics and experimental images illustrating the different interfacial regimes are shown in Figure~\ref{fig1} and Figure~\ref{fig2}(a), respectively. Remarkably, at moderate capillary numbers, when $Ca_2>1$ and $Ca_1\leq1$, the system exhibits a new regime termed the stream splitting regime, where the high-impedance liquid (HIL) continuously and simultaneously splits into droplets and a thin residual liquid stream along the channel wall (see Figure~\ref{fig2}(a)). As the capillary number increases further, for $Ca_2>1$ and $Ca_1>1$, the interface undergoes periodic undulations without droplet breakup, resulting in the interfacial waviness regime. At sufficiently high capillary numbers, for $Ca_2\gg1$ and $Ca_1\gg1$, a stable coflow is recovered. In contrast, reducing the capillary numbers to smaller values leads to a transition from splitting to stream relocation for $Ca_2\leq1$ and $Ca_1<1$, where the entire HIL stream relocates to the acoustic pressure node, and a further decrease results in complete stream-droplet breakup near the inlet for $Ca_2\ll1$ and $Ca_1\ll1$.

The transitions between these regimes are summarized in the regime map shown in Figure~\ref{fig2}(b), plotted in terms of the capillary numbers of the two liquid phases ($Ca_1$ and $Ca_2$) while keeping the acoustic input power constant at 125~mW. The gray, orange, green, purple and red shaded regions indicate the parameter space where stable coflow, waviness, splitting, relocation and stream-drop regimes are observed, while the yellow region denotes the no-coflow regime where coflow is not established even in the absence of ultrasound. The inset of Figure~\ref{fig2}(b) illustrates the shift of the splitting--waviness transition boundary with increasing acoustic input power. This dependence of acoustic power on regime transition will be discussed in more detail using a theoretical scaling model in \S~4.3.1.

To further understand the role of acoustic field, we recast the problem in terms of a characteristic length-scale ratio that compares the lateral position of the pressure nodal plane with the position of interface from the HIL-side wall. Specifically, we define the width ratio $W_r = W_{np}/W_1$, where $W_{np}$ is the distance of the pressure node from the HIL-side wall and $W_1$ is the width of the HIL stream, as illustrated schematically in Figure~\ref{fig1} and ~\ref{fig2}(c). The position of the acoustic node ($W_{np}$) is obtained by eigen-frequency analysis using simulations described in § \ref{sec:3.1}.

The variation of $W_r$ with the capillary number of the LIL stream for different regimes is shown in Figure~\ref{fig2}(c). This representation provides a complementary description of the flow regimes in terms of the relative position of the interface with respect to the pressure nodal plane. When $W_r<1$, the pressure node lies within the HIL stream, and the interface remains stable with only decaying perturbations. In contrast, when $W_r>1$, the pressure node lies within the LIL stream, such that the acoustic radiation force acts across the interface and drives the interface towards the nodal plane, leading to amplification of interfacial perturbations and the onset of acoustic instability. The specific regime that emerges is governed by the combined influence of $Ca_2$ and $W_r$. At low $Ca_2$ and $W_r>1$, the system exhibits the stream-droplet (SD) or relocation regime. As the capillary number increases and $W_r$ becomes slightly larger than unity, interfacial waviness develops. At even higher $W_r$, the interface undergoes splitting, producing droplets together with a thin residual stream (TRS) .

The observed interfacial behaviour also depends on the wetting characteristics of the HIL with the channel surface (see contact angle measurements in supplementary material S3). The fluid pairs exhibiting splitting and waviness show strong affinity toward the silicon walls (e.g., olive oil or mineral oil), with static contact angles $\theta_s \ll 90^\circ$ (see Table~\ref{tab:tab1}). Such wall-adherent liquids promote the formation of a thin liquid film along the wall, enabling the persistence of a thin residual stream during splitting. In contrast, for weakly wetting liquids with $\theta_s \sim 90^\circ$ (e.g., aqueous glycerol, presented in Appendix~\ref{appd}), splitting is not observed. Instead, such systems exhibit stream-to-droplet breakup for $Ca_{1,2}<1$ and $Ca_{ac}>1$, and relocation for $Ca_{ac}>1$ with $Ca_{1,2}>1$, consistent with \citep{Hemachandran2021}. Notably, stream-drop breakup and stream relocation have previously been reported \citep{Hemachandran2021}, and therefore, the present work primarily focuses on the unexplored splitting and waviness regimes due to their rich and complex interfacial dynamics. 

Finally, we observe that the transition between the splitting and waviness regimes is reversible. On-demand switching between these two regimes can be achieved by varying the capillary number at fixed acoustic power $P$ or by varying the acoustic input power while maintaining a constant capillary number, as demonstrated in Figure~\ref{fig2}(d). 

\subsection{Mechanism of acoustically induced splitting and waviness}\label{sec:4.3} 

In order to interpret the transition between stable, waviness, and splitting regimes, we compare the characteristic time scales governing interfacial deformation and downstream advection of the interface. Once the acoustic field initiates a perturbation at the interface, two competing timescales determine its subsequent evolution. The first is the pinching time scale associated with the growth of interfacial deformation driven by viscous stresses and opposed by interfacial tension. This time scale can be estimated as $t_p \approx \mu_1 U_1 W_1 / (\gamma_{12} U_2)$, which characterizes the time required for the interface to deform significantly under the combined influence of viscous forcing and capillary resistance. The second is the advection of the perturbation downstream by the co-flow, characterized by the advection time scale $t_{ad} \approx 2W / (U_1 + U_2)$, which represents the time required for the interface to be advected over a characteristic length scale. The ratio of these two time scales, $\chi= t_p/t_{ad}$, captures the competition between interfacial deformation and advective transport. When $\chi<0.5$, the deformation grows rapidly that pinching dominates over advection, leading to stream splitting regime with droplet formation accompanied by a thin residual stream. For intermediate values $0.5 < \chi < 1$, the perturbations are advected downstream before significant pinching occurs, producing periodic interfacial undulations, characteristic of the waviness regime. When $\chi> 1$, advection dominates over deformation, and the interface remains stable despite the presence of acoustic forcing.

\begin{figure}
\centering
\includegraphics[clip,width=1\textwidth]{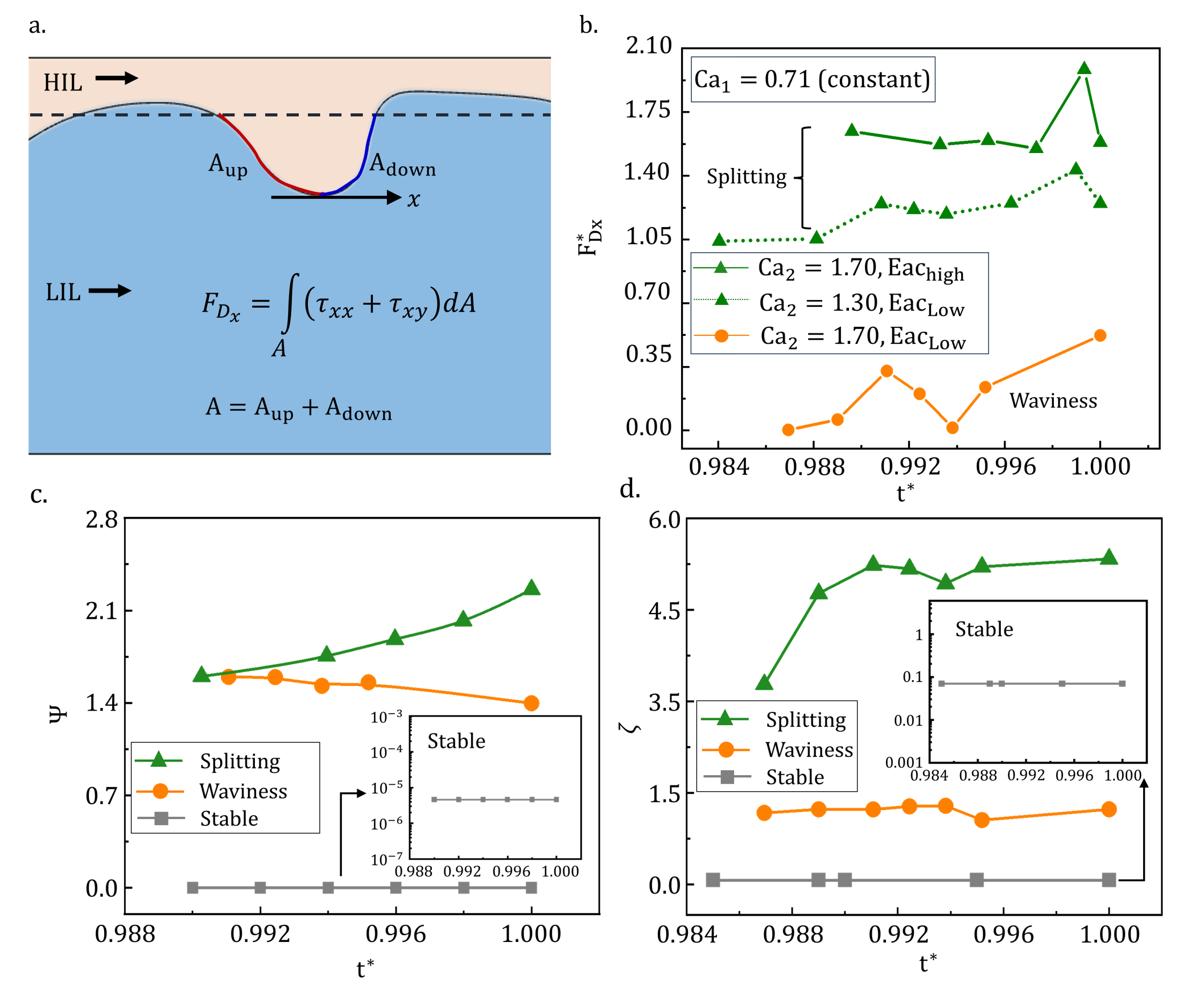}
\caption
{
\justifying{(a) Schematic of a liquid protrusion, resulting from interfacial deformation, the net drag force ($F_{D}$) acting on it is computed by integrating the stresses over the entire protruded surface area as $F_{D_x} = \int_{A} \left( \tau_{xx} + \tau_{xy} \right)\, dA$. The drag force acting on the upstream face of the protruded area ($F_{D_{up}}$) is also calculated. (b) The temporal variation of normalized net drag force in the stream-wise direction, ($F^*_{D_x}=F_{D_x}/F_{D_x,up}$) with dimensionless time ($t^*$) for different combinations of $Ca_2$=1.30 and 1.70, and $E_{ac}$=$40~\mathrm{J\,m^{-3}}$ and $70~\mathrm{J\,m^{-3}}$. Here, time is normalized by the breakup time in the splitting cases and by the time of maximum deformation in the waviness case. (c) Temporal variation of the streamwise force ratio $\psi = F_{D_x}/F_{I_x}$, where $F_{D_x}$ is the supporting viscous drag and $F_{I_x}$ is the resisting interfacial tension force, both along the flow direction. Results are shown for stream splitting: $Ca_1=0.71$, $Ca_2=1.70$, $E_{ac}=70\,\mathrm{J\,m^{-3}}$), waviness: $Ca_1=0.71$, $Ca_2=1.70$, $E_{ac}=40\,\mathrm{J\,m^{-3}}$), and stable coflow: $Ca_1=2.00$, $Ca_2=3.60$, $E_{ac}=40\,\mathrm{J\,m^{-3}}$. 
(d) Temporal variation of cross-stream force ratio $\zeta = (F_{ac}+F_{D_y})/F_{I_y}$, where $F_{ac}$ is the acoustic radiation force, $F_{D_y}$ is the transverse viscous drag, and $F_{I_y}$ is the interfacial tension force, both in the transverse direction. The relative magnitudes of $\psi$ and $\zeta$ distinguish stable, waviness, and splitting regimes.}}
\label{fig3}
\end{figure}

We now examine the physical mechanism underlying the acoustically induced interfacial deformation from force considerations. Upon activation of the acoustic field at resonance, a half-wave standing acoustic field is established across the channel width. Owing to the acoustic impedance mismatch between the two liquids, scattering of the acoustic field at the interface generates a radiation force acting normal to the interface from the HIL toward the LIL. This forcing leads to the development of periodic interfacial undulations along the flow direction. The observed wavelength of these undulations closely matches the acoustic wavelength within the microchannel, indicating that the interfacial modulation is governed by the imposed acoustic field. Details of the interfacial wavelength are discussed in \S\,4.7. Each interfacial protrusion formed by acoustic forcing acts as a local obstruction to the surrounding flow (see Figure~\ref{fig3}(a)). As the LIL flows past the protrusion, the viscous drag acting on the upstream face becomes larger than that on the downstream face. The magnitude of this asymmetry increases with the protrusion amplitude, which is governed by the acoustic forcing. The subsequent evolution of the instability is therefore determined by the competition between acoustic forcing, streamwise viscous drag, and the restoring interfacial tension. When the streamwise drag significantly exceeds the resisting interfacial tension, the protrusion elongates with a finite relative velocity along the flow direction. This deformation progressively thins the neck connecting it to the main HIL stream. The neck eventually pinches off downstream ($\chi < 0.5$), producing droplets while leaving behind a thin residual stream (TRS) attached to the channel wall (experimentally observed necking is shown in supplementary material S4). This corresponds to the splitting regime. In contrast, when the viscous drag remains comparable to the interfacial tension, the drag asymmetry is insufficient to drive further deformation. The undulations are then advected downstream without breakup, corresponding to the waviness regime ($0.5<\chi<1$). In the stable regime, perturbations are advected downstream faster than they can grow ($\chi>1$), or the acoustic forcing is insufficient to induce appreciable deformation. This corresponds to conditions of strong advection (high capillary number) or weak acoustic forcing, under which the interface remains essentially flat.

To verify this mechanism, we analyse the force distribution along the interface using dynamic two-phase simulations based on the phase-field formulation (see \S\,3.2). Simulations are performed over the experimentally explored parameter range, with acoustic forcing varied through the acoustic energy density $E_{ac}$, which represents the acoustic input power. A reference case corresponding to the waviness regime was simulated at higher capillary numbers $Ca_1=0.71$ and $Ca_2=1.70$ with lower acoustic energy, $E_{ac}=40\,\mathrm{J\,m^{-3}}$. Two other cases were considered to probe the transition toward splitting: (i) the capillary number of the LIL was reduced from $Ca_2=1.70$ to $Ca_2=1.30$ while keeping the same $E_{ac}$, and (ii) increasing $E_{ac}$ to $70\,\mathrm{J\,m^{-3}}$ at fixed capillary numbers. In both cases, the interface evolves toward droplet formation with a thin residual stream, consistent with experimental observations of the splitting regime. 

To quantify the drag asymmetry, we evaluated the streamwise (x-direction) viscous drag acting on the liquid protrusions in the waviness and splitting cases. The net drag force ($F_{D}$) is evaluated by integrating the normal and tagential stresses over the protruded surface area, $F_{p_x} = \int_{A} \left( \tau_{xx} + \tau_{xy} \right)\, dA$ and a higher net drag force suggests a higher drag asymmetry. The net drag force will have both stream-wise (or $x$) and cross-stream (or $y$) components (i.e., $F_{D_x}$ and $F_{D_y}$). The temporal variation of normalized net drag force in the stream-wise direction, ($F^*_{D_x}=F_{D_x}/F_{D_x,up}$) with dimensionless time ($t^*$) is presented in Figure~\ref{fig3}(b). Here, $F_{D_x,up}$ is the x-component of the drag force acting on the upstream face of the protrusion, and time is normalized by the breakup time for the splitting case and by the time of maximum deformation for the waviness case. As observed, in the waviness regime, a smaller value of $F^*_{D_x}\sim 0.1$ suggests that the drag contributions remain nearly symmetric. In contrast, the splitting regime exhibits strong drag asymmetry, with $\Delta F^*_{D_x}>1$. This indicates that the upstream drag dominates and drives elongation of the protrusion, ultimately leading to droplet breakup accompanied by a thin residual stream.

To further examine the coupled roles of acoustic forcing, viscous drag, and interfacial tension, we evaluated ratios of forces supporting and resisting interfacial deformation along both streamwise (x-direction) and transverse (y-direction) directions. The streamwise force ratio is defined as $\Psi = F_{D_x}/F_{I_x}$, where $F_{D_x}$ represents the supporting viscous drag acting along the flow direction and $F_{I_x}$ denotes the resisting interfacial tension force. The temporal evolution of $\Psi$ is shown in Figure~\ref{fig3}(c). Similarly, the transverse force balance was characterised by the parameter $\zeta = (F_{ac}+F_{D_y})/F_{I_y}$, where $F_{ac}$ is the acoustic radiation force, $F_{D_y}$ the transverse viscous drag, and $F_{I_y}$ the transverse component of the interfacial tension force. The evolution of $\zeta$ with normalized time (normalized similarly to that in Figure~\ref{fig3}(b)) is shown in Figure~\ref{fig3}(d). The three different regimes are illustrated in Figures~\ref{fig3}(c) and \ref{fig3}(d). In the splitting regime $\Psi>>1$ and $\zeta>>1$, indicating that the combined viscous and acoustic forces dominate the resisting interfacial tension and drive progressive elongation of the protrusions until breakup occurs. In the waviness regime, $\zeta>1$ remains greater than unity but nearly constant in time, while $\Psi\sim1$. Under these conditions the acoustic forcing is sufficient to deform the interface, but the streamwise drag remains insufficient to produce breakup, resulting in sustained periodic undulations. In the stable regime, $\Psi<<1$ and $\zeta<<1$ (see insets in Figures~\ref{fig3}(c) and \ref{fig3}(d)), indicating that the supporting forces are too small to overcome resisting force and therefore interface remains essentially undeformed.

The mechanism responsible for the different regimes differs fundamentally from classical spontaneous interfacial instabilities. To place the observed behaviour in context, we compare it with canonical instability mechanisms commonly encountered in coflow systems. The Saffman--Taylor instability arises when a less viscous fluid displaces a more viscous one under a pressure gradient in confined geometries; however, in the present system no cross-stream pressure gradient exists, and the interfacial deformation is instead driven by time-periodic acoustic radiation pressure. The Kelvin--Helmholtz instability, which originates from strong interfacial shear, is also unlikely because the velocity contrast between the two streams is insufficient to overcome surface tension. Yih-type instabilities associated with viscosity-stratified flows at low Reynolds number arise spontaneously due to viscosity contrast, typically in vertically stratified configurations, whereas the present interface remains stable in the absence of acoustic forcing. Similarly, Rayleigh--Taylor and Marangoni instabilities can be excluded, as the experiments are conducted under horizontal orientation without density-driven stratification and under nearly isothermal conditions without interfacial tension gradients (see transient temperature study in supplementary material S5). Although none of these classical mechanisms individually explains the observed dynamics, the present instability shares certain features. The acoustic radiation force periodically displaces the interface, generating protrusions reminiscent of Saffman--Taylor fingering, but driven by acoustic radiation pressure rather than a fluid pressure gradient. These protrusions are subsequently advected and elongated downstream by viscous shear and drag imposed by the co-flowing streams. As the neck of the elongated structures thins, capillary forces act to minimize surface energy, ultimately leading to droplet pinch-off in a manner analogous to Rayleigh--Plateau breakup. Thus, acoustic radiation pressure initiates interfacial displacement, viscous stresses govern the subsequent elongation, and interfacial tension controls the final breakup.

\subsection{Effect of acoustic power on regime transition and interfacial force balance
}\label{sec:4.4}

We find that regime transitions are sensitive to the acoustic radiation force, which in turn depends on the applied acoustic power. In particular, regime boundaries in Figure~\ref{fig2}(b) are not fixed, and variations in the actuation power can shift the transition boundaries between the different flow regimes. Experimental evidence in the inset of Figure~\ref{fig2}(b) (indicated by dotted lines) shows that by increasing the power from $P_1$ to $P_2$ shifts the demarcation line between splitting and waviness upward, thereby expanding the splitting regime. These transitions are typically reversible and can be tuned back and forth by adjusting the input acoustic power, keeping the flow rates fixed (see Figure~\ref{fig2}(d)). The experimental images in Figure~\ref{fig2}(d) illustrate that for the same hydrodynamic conditions or capillary numbers, the reversible transition between the waviness and stream splitting regimes can be observed simply by varying the applied peak-to-peak voltage, $V_{pp}$. This reversible transition also highlights that splitting can be achieved even at higher $Ca$ by increasing the acoustic power, and conversely, reducing power can transition a splitting case to waviness. The ability to generate droplets at elevated $Ca$ values broadens the operational window, marking a notable advancement in droplet microfluidics. The effect of applied acoustic power on the regime transition is illustrated in terms of the force ratio, $\Psi$ (see Figure~\ref{fig3}(c)). In dynamic simulation, at a fixed flow condition or capillary numbers, a coflow system exhibits interfacial waviness at lower power $(E_{ac}=40 J/m^3)$ marked by a smaller force ratio, but transitions to stream splitting at higher power $(E_{ac}=70 J/m^3)$ as indicated by a large force ratio (see Figure~\ref{fig3}(c)). These results suggest that higher acoustic power accelerates the growth of interfacial instabilities, enhancing the local drag on protruding wavelets and triggering droplet formation, thereby splitting the HIL stream. 

In order to theoretically predict the transition boundary between waviness and splitting regimes, we developed a force scaling model based on the balance of relevant forces acting on a control volume surrounding a typical interfacial protrusion (see Appendix~\ref{appe}). We scaled the different forces to estimate the position of the regime boundary without solving complex higher order coupled equations. The acoustic radiation force (see eqn.~\ref{Eqn-3.7}) acting on the control volume scales as $\mathbf{F}_{ac} \sim E_{ac}~ A_I~ \mathcal{C_e}~ \mathbf{n}$, where $E_{ac}$ is the acoustic energy density, $A_I$ is the effective interfacial area, $\mathbf{n}$ is the unit normal to the interface, and $\mathcal{C_e}$ is a dimensionless complex coefficient containing reflection and transmission contributions arising from impedance and density mismatch across the interface. The coefficient $\mathcal{C_e}$ contains exponential factors associated with the spatial phase of the standing wave, which do not affect the magnitude of the scaled force. Consequently, $|\mathcal{C_e}| \sim \mathcal{O}(\Phi)$, where $\Phi$ is the acoustic contrast factor. Because the net acoustic stress vanishes for acoustically matched fluids, the leading non-zero contribution must be proportional to material property mismatch. For liquid--liquid interfaces, the contrast factor scales with impedance mismatch as $\Phi \sim \overline{\Delta Z}$, where $\overline{\Delta Z} = (Z_1 - Z_2)/(Z_1+Z_2)$ and $Z_i = \rho_i c_i$ is the acoustic impedance of phase $i$ (reference). For a single wavelet, the effective interfacial area over which the radiation traction acts scales as $A_I \sim h \lambda_H$, where $h$ is the channel height and $\lambda_H$ is the characteristic wavelet length. The total acoustic force acting on a wavelet therefore scales as $F_{ac} \sim E_{ac} ~\overline{\Delta Z}~h \lambda_H$.

The interfacial restoring force is governed by the Young–Laplace relation, which states that the pressure jump across a curved interface is $\Delta p_\gamma = \gamma \kappa$, where $\gamma$ is the interfacial tension and $\kappa$ is the mean curvature. For a wavelet-like protrusion with characteristic streamwise length scale $\lambda_H$ and confinement height $h$, the principal radii of curvature scale as $R_1 \sim \lambda_H$ (streamwise direction) and $R_2 \sim h$ (cross-sectional confinement direction). The corresponding capillary pressure is thus $\Delta p_\gamma \sim \gamma (1/\lambda_H + 1/h)$. The net restoring force is obtained by multiplying this pressure by the projected interfacial area which scales as $A_I \sim h \lambda_H$. Consequently, the total interfacial tension force scales as $F_I \sim \gamma (1/\lambda_H + 1/h) (h \lambda_H)$.
Simplifying, we get $F_I \sim \gamma (h + \lambda_H)$. We assume the same scaling for the interfacial tension force in both the streamwise and cross-stream directions because capillarity enters the interfacial stress balance through the scalar Laplace pressure $\Delta p_\gamma = \gamma \kappa$, which generates a normal restoring stress determined solely by the local mean curvature. Since this normal capillary stress is isotropic, its order-of-magnitude contribution to restoring deformation is identical when projected onto the streamwise and cross-stream directions. 
 
The viscous drag forces in the stream-wise and cross-stream directions arise from the jump in the tangential shear stress across the interface. For a Newtonian liquid, the dominant interfacial shear stress scales as $\tau \sim \mu\,\partial U/\partial y$, since velocity gradients are primarily imposed by confinement across the channel width. The streamwise viscous force is therefore obtained by multiplying the shear stress difference between the two fluids by the effective interfacial area $A_x \sim h W_0$. The stream-wise drag force scales as $F_{D_x} \sim (\mu_2 \partial U_2/\partial y - \mu_1 \partial U_1/\partial y)\, h W_0$, where $U_1$ and $U_2$ denote the streamwise velocities of the HIL and LIL, respectively, $\mu_1$ and $\mu_2$ are corresponding dynamic viscosities, and $W_a$ is the wavelet amplitude, taken as one-third of the channel width from experimental observations. Similarly, cross-stream deformation induces transverse velocities $V_1$ and $V_2$. Owing to geometric confinement, velocity variations across the channel height dominate over streamwise variations, so that $\partial V/\partial y \gg \partial V/\partial x$. The strongest shear therefore develops across the confinement direction, leading to a dominant viscous stress scaling as $\tau \sim \mu\,\partial V/\partial y$. Multiplying the interfacial stress difference by the projected wavelet area in the streamwise--vertical plane, $A_I \sim h \lambda_H$, gives $F_{D_y} \sim (\mu_2 \partial V_2/\partial y - \mu_1 \partial V_1/\partial y)\, h \lambda_H$, where the transverse velocities are approximated as $V_i \sim (U_2 - U_1)/2$ following \cite{Hinch1984}. 
Resolving the forces in the $x$ direction (i.e., along the flow direction) as $F_{D_x}\sim F_{I_x}$, and assuming a linear variation of stress, we get
\begin{equation}\label{Eqn.4.1}
\lambda_H\sim (Ca_2  - Ca_1  - 2)h/2.	   
\end{equation}
Next, through force balance in the $y$ direction by equating the drag force to the net acoustic and interfacial forces as $F_{Dy}\approx F_{ac}  - F_{Iy}$, we obtain
\begin{equation}\label{Eqn.4.2}
\lambda_H\sim (F_{ac}/2\gamma- h )/(1+h/(4W_0 ) (1 - \mu_r )Ca_2  +h/(4W_0 ) (1 -1/\mu_r )Ca_1 ).		   
\end{equation}
By equating  Eqn.~\ref{Eqn.4.1} and  Eqn.~\ref{Eqn.4.2}, we obtain the condition for splitting to waviness transition as
\begin{multline}
K_1 K_2 (Ca_{1}^2 + \mu_r Ca_{2}^2) 
+ K_1 K_2 (K_2 - 2) Ca_{1} Ca_{2} \\
+ (2K_1 K_2 - \mu_r) Ca_{1} 
+ (1 - 2K_1 K_2)\mu_r Ca_{2}
\approx F_{ac}/\gamma h,
\end{multline}

\noindent where, $K_1=h/(4W_0)$ and $K_2=(1-\mu_r)$. The above analytical expression is plotted in the inset of Figure~\ref{fig2}(b) in terms of $Ca_1$ and $Ca_2$ to obtain the theoretically predicted boundary between the splitting and waviness regimes. While the scaling prediction captures the overall trend and the location of the transition boundary, a small difference is observed when compared with the experimental results. This deviation can be attributed to the simplifying assumptions inherent in the model, including the use of order-of-magnitude force scaling, neglect of higher-order geometric effects, and the assumption of linear stress variation and idealized interfacial shapes. Nevertheless, the model successfully captures the correct dependence of the transition boundary on the governing parameters and, importantly, predicts the shift in the regime boundary with increasing acoustic power. In particular, higher acoustic power shifts the transition toward higher capillary numbers, thereby expanding the splitting regime, consistent with experimental observations. The agreement between model predictions and experimental results demonstrates that the transition dynamics is governed by the coupled interplay between acoustic radiation force, viscous drag, and interfacial tension, with the acoustic forcing acting as the key control parameter.

\subsection{Breakup length in splitting regime}\label{sec:4.5}

\begin{figure}
\centering
\includegraphics[clip,width=0.8\textwidth]{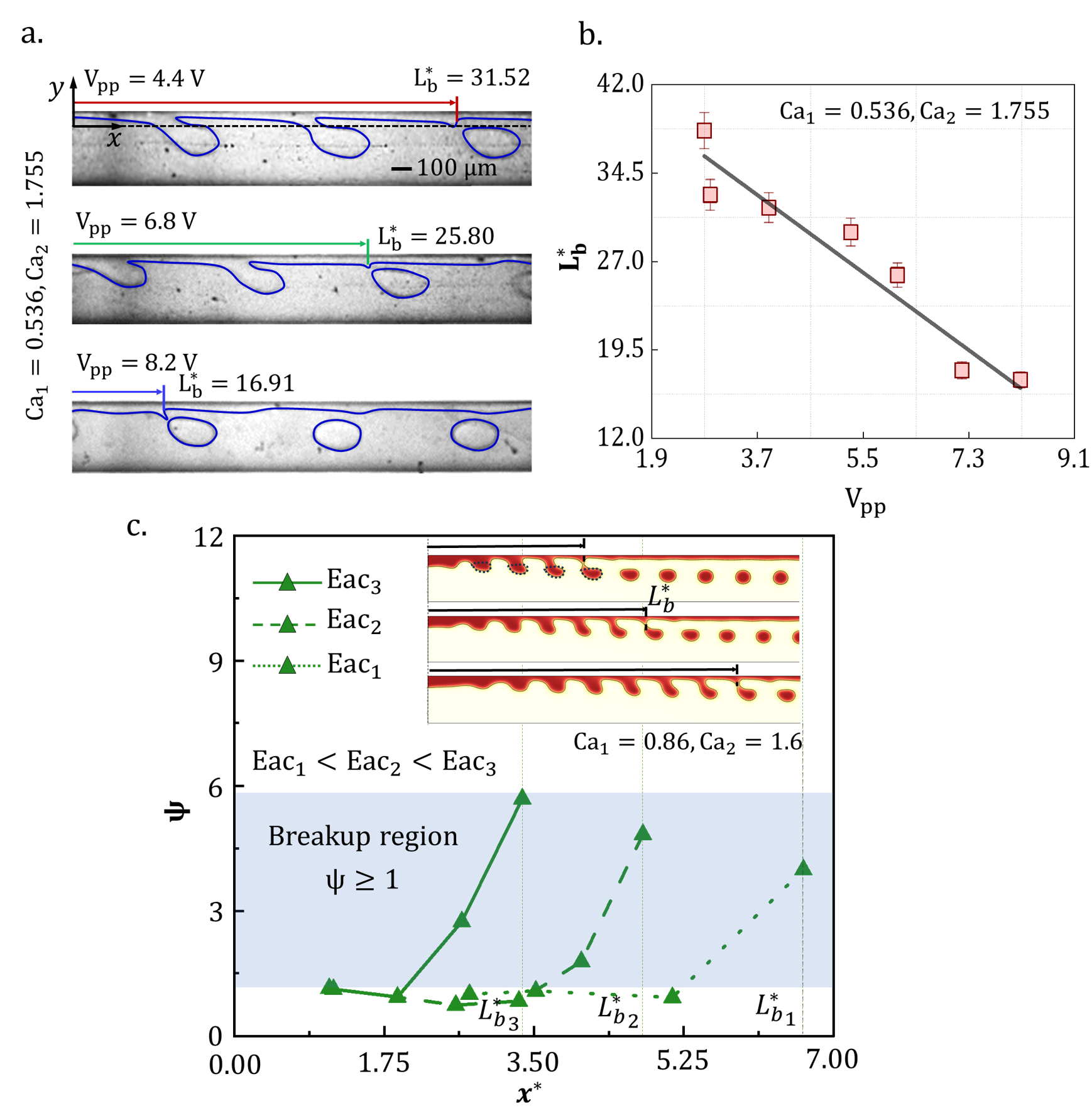}
\caption
{
\justifying{(a) Experimental images illustrating the influence of acoustic excitation, expressed in terms of the peak-to-peak voltage $V_{pp}$, on the droplet breakup location along the channel. The breakup length $L_b$ decreases as the acoustic power increases. (b) Variation of the normalized breakup length $L_b^*$ ($=L_b/W$) with applied peak-to-peak voltage $V_{pp}$ at fixed capillary numbers. In (a) and (b): the capillary numbers are fixed at $Ca_1 = 0.536$ and $Ca_2 = 1.755$. Experimental snaps correspond to the Silicon-Olive oil coflow system. Scale bar: $100\,\mu$m. (c) Numerical prediction of $L_b$ from the variation of the streamwise force ratio $\Psi = F_{D_x}/F_{I_x}$ along the flow direction $x^*$ ($=x/W$) for different acoustic energy densities, $E_{ac_1} < E_{ac_2} < E_{ac_3}$. The position where $\Psi\gtrsim1$ corresponds to the onset of droplet pinch-off and therefore predicts the breakup location $L_b^*$. Increasing acoustic forcing shifts this location toward the inlet, consistent with the experimental observations in (a) and (b). The inset shows simulation snapshots illustrating the corresponding breakup locations. The capillary numbers are fixed at $Ca_1 = 0.86$ and $Ca_2 = 1.6$.}}
\label{fig4}
\end{figure}

We study how acoustic forcing and hydrodynamic parameters control the characteristics of droplet breakup in the splitting regime. First, we analyze the breakup length $L_b$, defined as the streamwise distance from the inlet junction (marked IJ in Figure~\ref{fig1}) to the location of droplet pinch-off. Breakup length normalized by the channel width $L_b^* = L_b/W$, serves as a key indicator of the growth rate of acoustically induced interfacial perturbations. A shorter breakup length corresponds to faster amplification of the instability, whereas a longer breakup length indicates slower growth and delayed droplet formation. Notably, in the absence of acoustic forcing, the interface remains hydrodynamically stable, highlighting the role of acoustics in triggering the instability. To investigate the effect of acoustic power $P$ on the breakup location, we conducted experiments at fixed flow conditions (i.e., fixed capillary numbers) while systematically varying the acoustic input power. Experimental images illustrating the variation of $L_b$ under different acoustic excitation levels are shown in Figure~\ref{fig4}(a). The experiments were performed at fixed capillary numbers $Ca_1 = 0.536$ and $Ca_2 = 1.755$, while the acoustic power was varied through the peak-to-peak voltage applied to the transducer ($V_{pp} = 4.4\,\mathrm{V}$, $6.8\,\mathrm{V}$, and $8.2\,\mathrm{V}$). As the acoustic power increases, the breakup location shifts progressively toward the inlet, indicating that stronger acoustic forcing accelerates the growth of interfacial instabilities.

The variation in the normalized breakup length $L_b^*$ with the applied voltage $V_{pp}$ is presented in Figure~\ref{fig4}(b). The experimental data show an approximately linear decrease in $L_b^*$ with increasing acoustic power, with $R^2 = 0.916$. The results suggest that the breakup location can be spatially controlled through modulation of the acoustic input power. A higher acoustic power enhances the acoustic radiation force, increasing the interface deformation and thus strengthening the viscous drag imbalance on the interfacial protrusions. Consequently, the condition for breakup is satisfied upstream, resulting in early splitting and droplet formation.

To complement the experimental observations and elucidate the underlying mechanism, numerical simulations were performed to analyse the spatial evolution of the interfacial force balance. In particular, we evaluated the streamwise force ratio $\Psi = F_{D_x}/F_{I_x}$ at successive protrusions formed along the interface at different streamwise locations, where $F_{D_x}$ represents the destabilizing viscous drag force and $F_{I_x}$ denotes the resisting interfacial tension force. The variation of $\Psi$ with the dimensionless streamwise coordinate $x^* = x/W$ for three increasing levels of acoustic energy density, $E_{ac_1}<E_{ac_2}<E_{ac_3}$ is shown in Figure~\ref{fig4}(c). The inset presents representative simulation snapshots indicating the protrusions used for the force integration, where the last necking point in each case corresponds to the droplet pinch-off location. These predicted breakup locations are denoted by $L_{b_1}^*$, $L_{b_2}^*$, and $L_{b_3}^*$ for $E_{ac_1}$, $E_{ac_2}$, and $E_{ac_3}$, respectively. Consistent with the experimental observations in Figure~\ref{fig4} (a) and (b), the simulations show that increasing acoustic forcing shifts the breakup location upstream, such that $L_{b_3}^*<L_{b_2}^*<L_{b_1}^*$. The results show that stronger acoustic forcing increases the viscous drag acting on the interfacial protrusions relative to the resisting interfacial tension, thereby allowing the breakup condition to be satisfied earlier along the channel. Consequently, the breakup length decreases with increasing acoustic power.

We investigate the influence of the combined flow rate and viscosity ratio parameter $Q_r \mu_r$ on the breakup location, at a fixed acoustic power. Here, $Q_r = Q_1/Q_2$ is the flow rate ratio and $\mu_r = \mu_1/\mu_2$ is the viscosity ratio between the high-impedance liquid (HIL) and the low-impedance liquid (LIL). All experiments were conducted at a fixed acoustic input power so that the acoustic forcing remained constant while only the hydrodynamic conditions were varied. The corresponding data are presented in Appendix~\ref{appf}.

Unlike the systematic behavior observed with acoustic power, the breakup length does not exhibit a clear monotonic dependence on $Q_r\mu_r$. Instead, the data show significant scatter, indicating that the breakup location varies irregularly with changes in the hydrodynamic parameters. This behaviour reflects the complex coupling between viscous stresses and interfacial tension under confined coflow conditions. As discussed in the mechanism section, droplet pinch-off is governed by the balance between the viscous drag acting on interfacial protrusions and the resisting interfacial tension. Variations in $Q_r$ and $\mu_r$ modify the local velocity field and viscous stresses at the interface in a nonlinearly, leading to spatially varying deformation dynamics along the channel. Consequently, the location where the breakup condition is satisfied, does not shift in a consistent or predictable manner when only hydrodynamic parameters are varied. Thus, compared with hydrodynamic parameters such as flow-rate or viscosity ratios, acoustic power provides a significantly more robust and predictable means of controlling the spatial location of droplet formation in the splitting regime.

\subsection{Interfacial oscillation in splitting and waviness regimes}\label{sec:4.6}

\begin{figure}
\centering
\includegraphics[clip,width=1\textwidth]{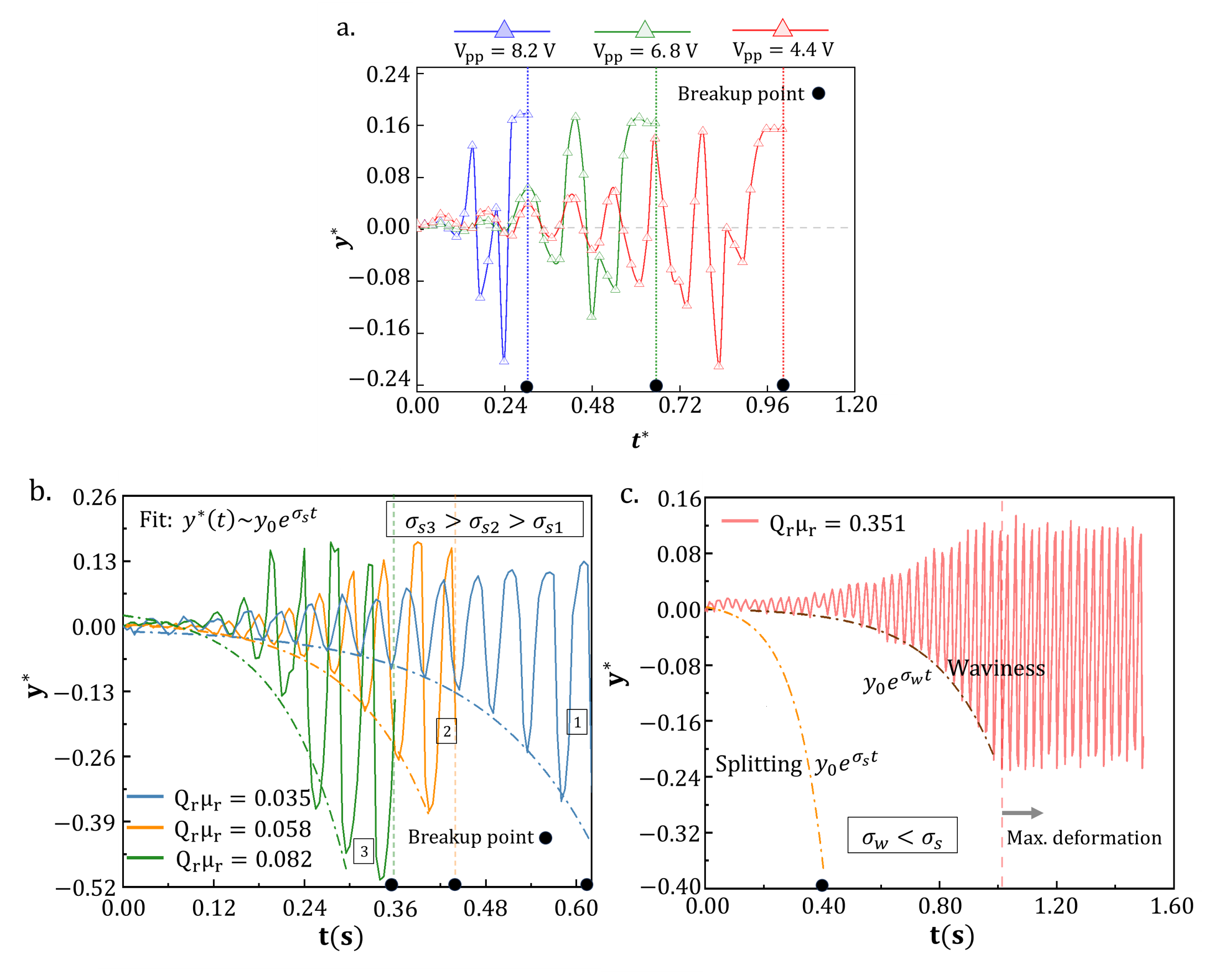}
\caption
{
\justifying{ (a) Temporal evolution of the interfacial oscillation amplitude $y^*$ measured at the droplet pinch-off location for different values of $V_{pp}$ at fixed capillary number. The capillary numbers are fixed at $Ca_1 = 0.536$ and $Ca_2 = 1.755$. (b) and (c): Experimental results showing the temporal evolution of the interfacial oscillation amplitude $y^*$ under identical acoustic power. (b) \textit{Splitting regime:} time evolution of the interfacial displacement for three representative values of $Q_r \mu_r$, where $Q_r = Q_1/Q_2$ is the flow-rate ratio and $\mu_r = \mu_1/\mu_2$ is the viscosity ratio. Solid curves represent experimentally measured interfacial oscillations, while dashed lines denote exponential envelope fits obtained from peak amplitudes using $y^*(t) \sim y_0 \exp(\sigma_s t)$. Within the splitting regime, increasing $Q_r \mu_r$ leads to a higher growth rate $\sigma_s$, resulting in rapid amplification of perturbations, interfacial necking, and eventual pinch-off (breakup). (c) \textit{Waviness regime:} temporal evolution of interfacial oscillations for a representative case, showing bounded oscillations without breakup under the same acoustic power. The exponential envelope $y^*(t) \sim y_0 \exp(\sigma_w t)$ indicates a lower growth rate $\sigma_w$ compared to the splitting regime, resulting in sustained but non-breaking oscillations. For comparison, the growth behaviour associated with the splitting regime is also indicated, highlighting the contrast in amplification dynamics. These results demonstrate that the transition between waviness and splitting is governed by the magnitude of the temporal growth rates, with $\sigma_s > \sigma_w$ leading to instability and breakup, while smaller growth rates result in bounded oscillatory behaviour.}}
\label{fig5}
\end{figure}

To understand the temporal evolution of instability in the splitting regime, we measured the interface displacement $y(t)$ perpendicular to the flow direction at the droplet breakup region. The interfacial displacement was measured from high-speed experimental recordings by tracking the position of the interface at a fixed streamwise location corresponding to the necking region where droplet formation eventually occurs. For each experimental condition, the interface position was extracted frame-by-frame using image processing, and the displacement was defined as the transverse deviation of the interface from its initial flat position prior to acoustic actuation. The temporal evolution of $y(t)$ was then obtained starting from the instant when the acoustic field was applied. The time evolution of the interfacial oscillations at the droplet breakup points for different acoustic power levels is shown in Figure~\ref{fig5}(a). The measurements are taken from the instant of acoustic field application until the formation of the first droplet. The results show that increasing acoustic power leads to faster growth of oscillation amplitude, suggesting enhanced instability growth rate. As a consequence, the breakup location shifts upstream with increasing acoustic power. Notably, the oscillation amplitude just prior to breakup remains comparable across different acoustic power levels. This suggests that the critical deformation required for droplet pinch-off is largely independent of the applied acoustic power, implying that the droplet size is not directly controlled by the strength of the acoustic force, as experimentally demonstrated in \S~4.8.1.

We analyze the time evolution of the interfacial displacement $y(t)$ for the splitting regime (Figure \ref{fig5}(b)) and the waviness regime (Figure \ref{fig5}(c)) at fixed acoustic power but different $Q_r\mu_r$. In the splitting regime, three representative cases are shown that correspond to increasing values of the dimensionless parameter $Q_r\mu_r$. For each case, the oscillation amplitude $y(t)$ is tracked until the formation of the first droplet, marking the transition from interfacial deformation to pinch-off. In contrast, the waviness regime exhibits sustained oscillations that remain bounded in amplitude and do not lead to droplet formation. After an initial growth phase, the oscillations reach a finite-amplitude oscillatory state, indicating that the destabilizing forces are insufficient to overcome the restoring interfacial tension.

To quantitatively characterize the instability dynamics, the temporal evolution of the interfacial displacement was fitted using an exponential growth model of the form $y(t) \sim y_0 e^{\sigma t}$, where $y_0$ represents the initial perturbation amplitude and $\sigma$ denotes the temporal growth rate of the instability. The fitted curves shown in Figure~\ref{fig5}(b) and (c) represent exponential envelope fits obtained from peak amplitudes using least-squares fitting.

In the splitting regime in Figure~\ref{fig5}(b), the fitted initial amplitudes lie in the range $y_0 \approx 0.44$--$0.69$, while the growth rates are $\sigma_s \approx 0.025$--$0.032$. In contrast, the waviness regime in Figure~\ref{fig5}(c) exhibits significantly smaller perturbations with $y_0 \approx 0.11$ and a lower growth rate $\sigma_w \approx 0.020$. These differences indicate that interfacial disturbances in the splitting regime both initiate with larger amplitudes and grow more rapidly than those in the waviness regime, leading to progressive amplification of the interface deformation until droplet pinch-off occurs. A comparison among the splitting cases further reveals a systematic dependence of the growth rate on the hydrodynamic parameter $Q_r\mu_r$. As $Q_r\mu_r$ increases from $0.035$ to $0.082$, the fit growth rate $\sigma_s$ increases, indicating that interfacial disturbances amplify more rapidly at higher values of $Q_r\mu_r$, leading to faster progression toward pinch-off. In contrast, the waviness regime is characterized by a relatively smaller initial amplitude $y_0$ and growth rate $\sigma_w$, which is insufficient to drive the system to breakup. Instead, the interfacial oscillations remain bounded. Overall, this analysis highlights that the transition between waviness and splitting is governed not only by the presence of acoustic forcing but also by the temporal growth characteristics of the interfacial perturbations, with $\sigma_s > \sigma_w$ leading to instability and breakup, while smaller growth rates result in bounded oscillatory behaviour.
\subsection{Variation of interfacial wavelength in splitting and waviness regimes}\label{sec:4.7}

\begin{figure}
\centering
\includegraphics[clip,width=1\textwidth]{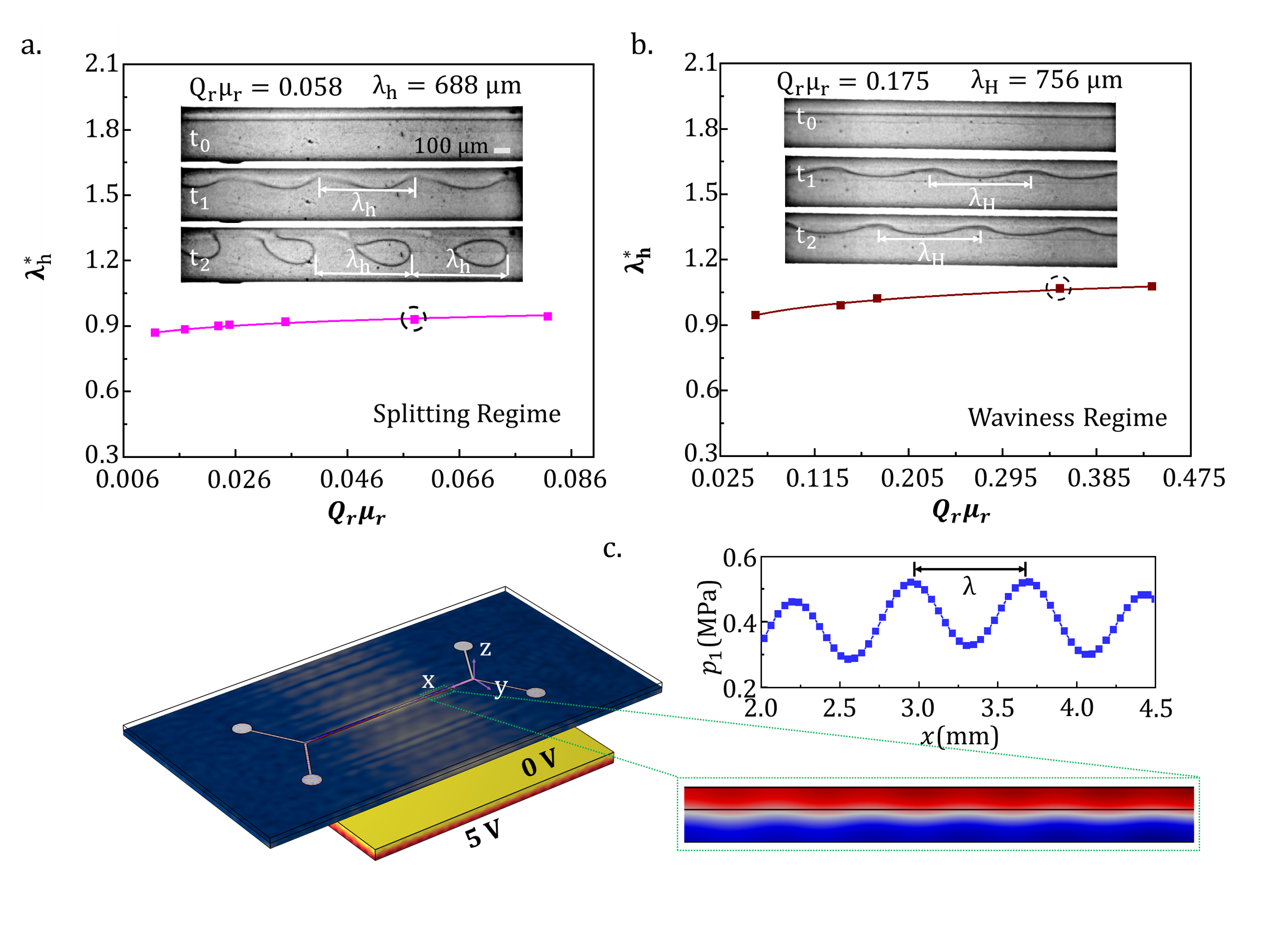}
\caption
{
\justifying{Variation of normalized interfacial wavelength ($\lambda_h^*=\lambda_h/\lambda$) with $Q_r \mu_r$ in (a) Splitting and (b) Waviness regimes. Insets show representative experimental snapshots corresponding to the marked data points. Here $t_0$ denotes the initial time, while $t_1$ and $t_2$ represent subsequent time instances, demonstrating that the interfacial wavelength remains nearly constant during the temporal evolution in both regimes. Experimental snapshots correspond to the silicon--olive oil coflow system.}
}
\label{fig6}
\end{figure}

We study spatial characteristics of the interfacial instability in the splitting and waviness regimes by experimentally analysing the wavelength of the resulting interfacial modulation. Remarkably, we find that the interfacial wavelength closely match with the wavelength of the ultrasound actuation. To quantify this, we define a non-dimensional wavelength parameter $\lambda_h^* = \lambda_h / \lambda$, where $\lambda_h$ is the experimentally measured interfacial wavelength and $\lambda$ is the acoustic wavelength inside the microchannel ($\lambda \approx 740~\mu\mathrm{m}$). The variation of $\lambda_h^*$ as a function of the hydrodynamic parameter $Q_r \mu_r$ for both splitting and waviness regimes is shown in Figure~\ref{fig6}(a) and (b). The wavelength $\lambda_h$ is obtained from experimental images by measuring the distance between successive peaks of the interfacial deformation along the stream-wise direction and is averaged over multiple realizations to minimize uncertainty. Remarkably, the results show that $\lambda_h^*$ remains close to unity throughout the entire range of $Q_r \mu_r$, confirming that the interfacial wavelength closely follows the imposed acoustic wavelength over a range of hydrodynamic conditions. The experimental data can be approximated by a fitted relation $\lambda_h^* \approx (Q_r \mu_r)^{0.05 \pm 0.01}$, indicating a very weak dependence of the interfacial wavelength on the hydrodynamic conditions. In our experiments, $Q_r \mu_r < 1$, and therefore the exponent of the fit remains small, resulting in only minor deviations of $\lambda_h^*$ from unity over varying hydrodynamic conditions.

To explain the close match between wavelengths of interfacial modulation and ultrasound actuation, we perform fully-coupled electromechanical simulations (discussed in \S\,3.3 and detailed in Appendix~G) to predict the formation of standing pressure fields within the microchannel. This standing-wave field establishes a spatially periodic distribution of the first-order acoustic pressure along the microchannel, with a well-defined nodal structure. The acoustic pressure ontained from simulations exhibits a periodic variation along the streamwise direction with a wavelength consistent with the imposed acoustic wavelength. This agreement demonstrates that the spatial structure of the instability is dictated primarily by the externally imposed acoustic field rather than by the hydrodynamic parameters of the coflow. Thus, while the parameter $Q_r \mu_r$ influences the temporal growth and nonlinear evolution of the instability, the interfacial wavelength itself is set by the acoustic forcing, as consistently observed in both experiments and electromechanical simulations.

\subsection{Droplet size and thin residual stream thickness in splitting regime}\label{sec:4.8}

We study the geometric outcome of the splitting regime in terms of the droplet size and thickness of the thin residual stream (TRS). These two quantities describe how the volume of the high-impedance liquid (HIL) is partitioned between the detached droplets and the wall-adhered residual liquid layer during the acoustically induced splitting process.

\begin{figure}
\centering
\includegraphics[clip,width=0.9\textwidth]{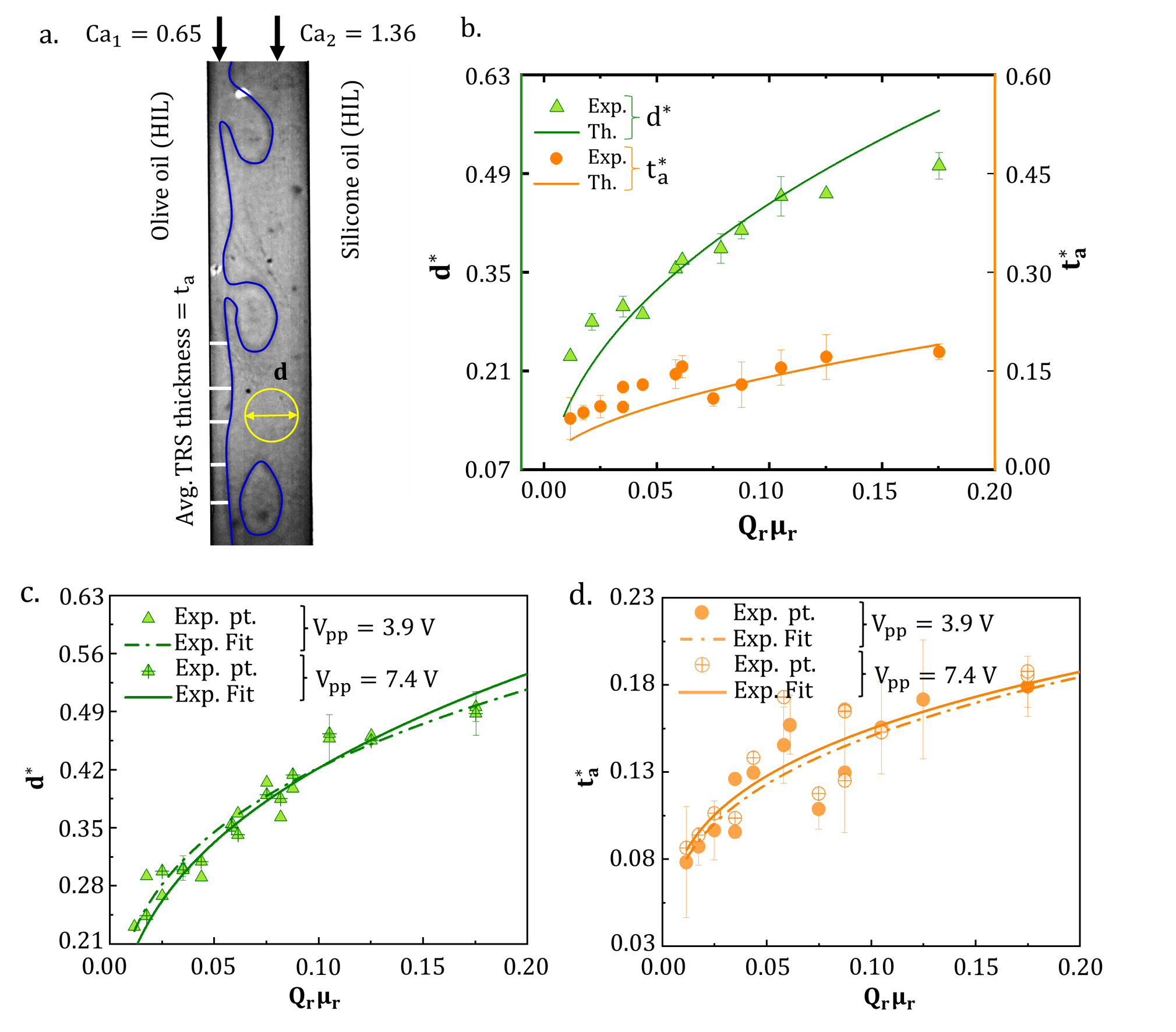}
\caption
{
\justifying{Characterization of droplet size and thin residual stream (TRS) thickness in the splitting regime. 
(a) Experimental image for silicone--olive oil pair, showing droplet fromation from HIL together with TRS along the channel wall. The droplet diameter $d$ is obtained from the equivalent spherical volume, while the TRS thickness $t_a$ is measured as the average distance between the channel wall and the interface (indicated by arrows) over multiple streamwise locations.(b) Variation of normalized droplet diameter $d^*=d/W$ (left axis, green) and normalized residual layer thickness $t_a^*=t_a/W$ (right axis, orange) as a function of $Q_r\mu_r$. Symbols represent experimental measurements, where green triangular markers correspond to droplet size and orange circular markers correspond to TRS thickness. Solid lines denote theoretical predictions obtained from the scaling relations. The monotonic increase of both $d^*$ and $t_a^*$ with $Q_r\mu_r$ reflects enhanced transport of HIL into interfacial protrusions prior to pinch-off.
}
}
\label{fig7}
\end{figure}

The droplet diameter is obtained from the captured high-speed experimental videos of droplet formation in the splitting regime, by extracting the projected area of each droplet using image analysis. Owing to the confinement along the channel height (\(h = 100~\mu\mathrm{m}\)), the droplets are not spherical; therefore, their volume was estimated as \(V_d = A_p\,h\), where \(A_p\) is the measured projected area. An equivalent spherical diameter was then obtained as \(d = (6V_d/\pi)^{1/3}\) (see Figure~\ref{fig7}(a)). The normalized droplet size is then defined as \(d^* = d/W\), where \(W\) is the channel width. 

The thickness of the thin residual stream \(t_a\) was measured from experimental images as the distance between the channel wall and the liquid--liquid interface after droplet detachment. For each case, a streamwise region of length close to one wavelength (\(\lambda_h\)) was considered and divided into five equally spaced locations (see Figure~\ref{fig7}(a)). The local liquid film thickness at each location was tracked over time using post-breakup frames, and a time-averaged thickness was obtained at each position. The final value of \(t_a\) was then determined by averaging these five time-averaged measurements, providing a representative thickness for a given flow condition in the splitting regime. The normalized thickness is defined as \(t_a^* = t_a/W\).

The experimentally measured variation of \(d^*\) and \(t_a^*\) as functions of the hydrodynamic parameters, in terms of \(Q_r \mu_r\), are presented in Figure~\ref{fig7}(b). Each data point in figure~\ref{fig7}(b) corresponds to an independent experimental measurement obtained under steady splitting conditions, while figure~\ref{fig7}(a) shows a representative experimental frame illustrating both the detached droplet and the residual thin layer used for the measurements. The measurements show that both the normalized droplet diameter \(d^*\) and the normalized thin-layer thickness \(t_a^*\) increase monotonically with \(Q_r\mu_r\). Physically, this trend reflects the increasing supply of HIL to the growing interfacial protrusions as the flow-rate ratio increases. Larger values of \(Q_r\mu_r\) therefore deliver a greater volume of HIL to each protrusion before pinch-off occurs, producing larger droplets and leaving behind a thicker residual layer along the wall. Conversely, at smaller \(Q_r\mu_r\), the amount of HIL transferred to each protrusion is reduced, resulting in smaller droplets and a thinner residual stream.

To study the effect of acoustic forcing on the splitting outcome, we compare in Figure~\ref{fig7} the experimentally measured droplet size and thin residual layer thickness at two acoustic excitation levels, \(V_{pp}=3.9\,\mathrm{V}\) and \(V_{pp}=7.4\,\mathrm{V}\). The variation of the normalized droplet diameter \(d^*\) with \(Q_r\mu_r\) is presented in Figure~\ref{fig7}(c), while the corresponding measurements for the normalized thin-layer thickness \(t_a^*\) is presented in Figure~\ref{fig7}(d).

In both cases, the data obtained at the two acoustic excitation levels collapse onto nearly identical curves, indicating that the droplet size and residual-layer thickness remain essentially unchanged over the examined range of acoustic actuation. This demonstrates that once the splitting regime is triggered, the final droplet size and TRS thickness are largely insensitive to the magnitude of the applied acoustic power. This is further supported by a theoretical model presented below (see Eqs.~\ref{eqn4.4} and~\ref{eqn4.6}), which shows no explicit dependence of these two quantities on acoustic forcing, confirming that the final breakup characteristics are governed primarily by viscous–capillary dynamics.

To interpret these measurements quantitatively, we develop a scaling framework to predict the droplet size and residual-layer thickness. Importantly, the solid curves shown in figure~\ref{fig7}(c) and (d) are not empirical fits; rather, they represent theoretical predictions obtained directly from the scaling relations using given experimental conditions in terms of material properties and geometric parameters.

The normalized droplet size is predicted by considering a local viscous–capillary force balance governing droplet detachment. In contrast to the global force balance used for regime demarcation, the present scaling is based on forces acting at the scale of the interfacial protrusion that evolves into a droplet. The dominant forces are the streamwise viscous drag exerted by the co-flowing LIL and the resisting interfacial tension associated with the curvature of the protrusion. The viscous force can be estimated as $F_v\sim \mu_2 U_2 \pi d^2/(4W_1)$, where \(d\) is the characteristic size of the protrusion and \(W_1\) is the width of the HIL stream. The resisting capillary force scales as \(F_c \sim \gamma \pi d\), reflecting the curvature of the interface at the droplet scale. Balancing these forces yields a scaling relation for the droplet size, from which the normalized droplet diameter is obtained as
\begin{equation}\label{eqn4.4}
d^* = \frac{4\gamma h W}{\mu_2 Q_2}\left[\beta^{-2} - \beta^{-4}\right],
\end{equation}
where
\begin{equation}
\beta = \left(1 + 0.3(Q_r\mu_r)^{1/3}\right)^2.
\end{equation}

For a given experimental condition, all parameters entering this relation are independently known. The viscosity ratio \(\mu_r\) is determined from the measured viscosities of the two fluids, while the flow-rate ratio \(Q_r\) is imposed experimentally. The interfacial tension \(\gamma\), channel height \(h\), and channel width \(W\) are measured independently. Therefore, once the operating condition is specified, the droplet size predicted by the scaling relation can be computed directly.

The thickness of the residual thin layer is estimated by enforcing conservation of the HIL volume during the splitting process. The incoming HIL is partitioned between the detached droplet and the residual layer remaining along the channel wall. This volume balance yields the normalized residual-layer thickness as (see detailed derivation shown in Appendix~\ref{appH})

\begin{equation}\label{eqn4.6}
t_a^* = \alpha \left(\frac{Q_r}{\beta} - 
\frac{\pi d^2|\beta - 1 - Q_r|}{8\lambda W Q_r}\right),
\end{equation}
with
\begin{equation}
\alpha = \frac{\beta - 1}{Q_r + \beta - 1}.
\end{equation}

Here \(\lambda\) is the wavelength of the interfacial modulation, which remains close to the imposed acoustic wavelength. Once $Q_r$, $\mu_r$, $d$, $\lambda$, and the channel geometry are known, $t_a^*$ can be evaluated directly from the above relation.

The agreement between the predicted trends and the experimental measurements indicates that the scaling framework captures the dominant physics governing both droplet formation and thin-layer development. Minor deviations between theory and experiment can arise from simplifications in the scaling analysis, including assumptions regarding droplet geometry, uniform wavelength, and the neglect of transient interfacial shape variations during pinch-off.

These results demonstrate that, although the acoustic field initiates the interfacial modulation that leads to splitting, the final droplet size and residual-layer thickness are primarily determined by hydrodynamic transport and capillary resistance. In other words, the acoustic field acts mainly as a trigger for the instability, while the subsequent partitioning of the HIL into droplets and residual layer is governed by viscous–capillary dynamics and mass conservation.

\section{Conclusion}\label{sec:5}

We studied the interfacial dynamics of an otherwise stable liquid–liquid coflow under the action of an externally imposed standing acoustic field in a rectangular microchannel. By systematically probing the interplay between acoustic forcing, viscous stresses, and interfacial tension, this study reveals a rich hierarchy of flow responses, spanning stable coflow, interfacial waviness, stream relocation, stream splitting and breakup. Among these, two previously unexplored regimes are investigated in detail. The waviness regime reflects a dynamic yet non-destructive state in which interfacial perturbations are sustained and advected downstream, highlighting a delicate balance between acoustic excitation and convective transport. More significantly, we uncover an entirely new splitting regime, in which a continuous high-impedance liquid stream undergoes periodic breakup into discrete droplets while simultaneously retaining a thin residual layer along the channel wall. This dual outcome, droplet generation coupled with wall-bound film formation, represents a departure from classical microfluidic breakup mechanisms. Strikingly, this acoustically induced breakup occurs at moderate to high capillary numbers, $Ca \gtrsim O(1)$, a regime where conventional viscous–capillary instabilities would typically inhibit droplet formation. The acoustic field thus acts as an external destabilizing agent, enabling access to breakup pathways that are otherwise inaccessible. Equally important is the degree of control afforded by this mechanism: the location of droplet formation can be tuned continuously by adjusting the acoustic power, allowing spatially programmable breakup within a simple straight channel. Furthermore, because the base coflow remains stable in the absence of excitation, droplet generation can be switched on and off on demand, providing a robust and reversible actuation strategy. The splitting regime also introduces an additional level of control through the residual liquid layer. Its thickness, along with the droplet size, is governed primarily by the combined parameter $Q_r\mu_r$, underscoring the role of viscous transport in setting the partitioning between the droplet and film phases. This provides a unified framework to tailor both dispersed and continuous fluid fractions. Together, these findings establish acoustic excitation as a powerful and versatile tool for inducing and controlling interfacial instabilities in confined coflow systems. The ability to achieve on-demand droplet generation, spatially tunable breakup, and simultaneous thin-film deposition within a geometrically simple platform opens new directions for acoustofluidic control of multiphase flows. Such capabilities hold promise for applications in controlled fluid partitioning, localized delivery, and patterned liquid deposition in next-generation microfluidic and lab-on-a-chip technologies.


\appendix

\renewcommand{\thefigure}{A\arabic{figure}}
\setcounter{figure}{0}
\section{Geometry of the computational Models}\label{appA}

\begin{figure}
\centering
\includegraphics[clip,width=0.9\textwidth]{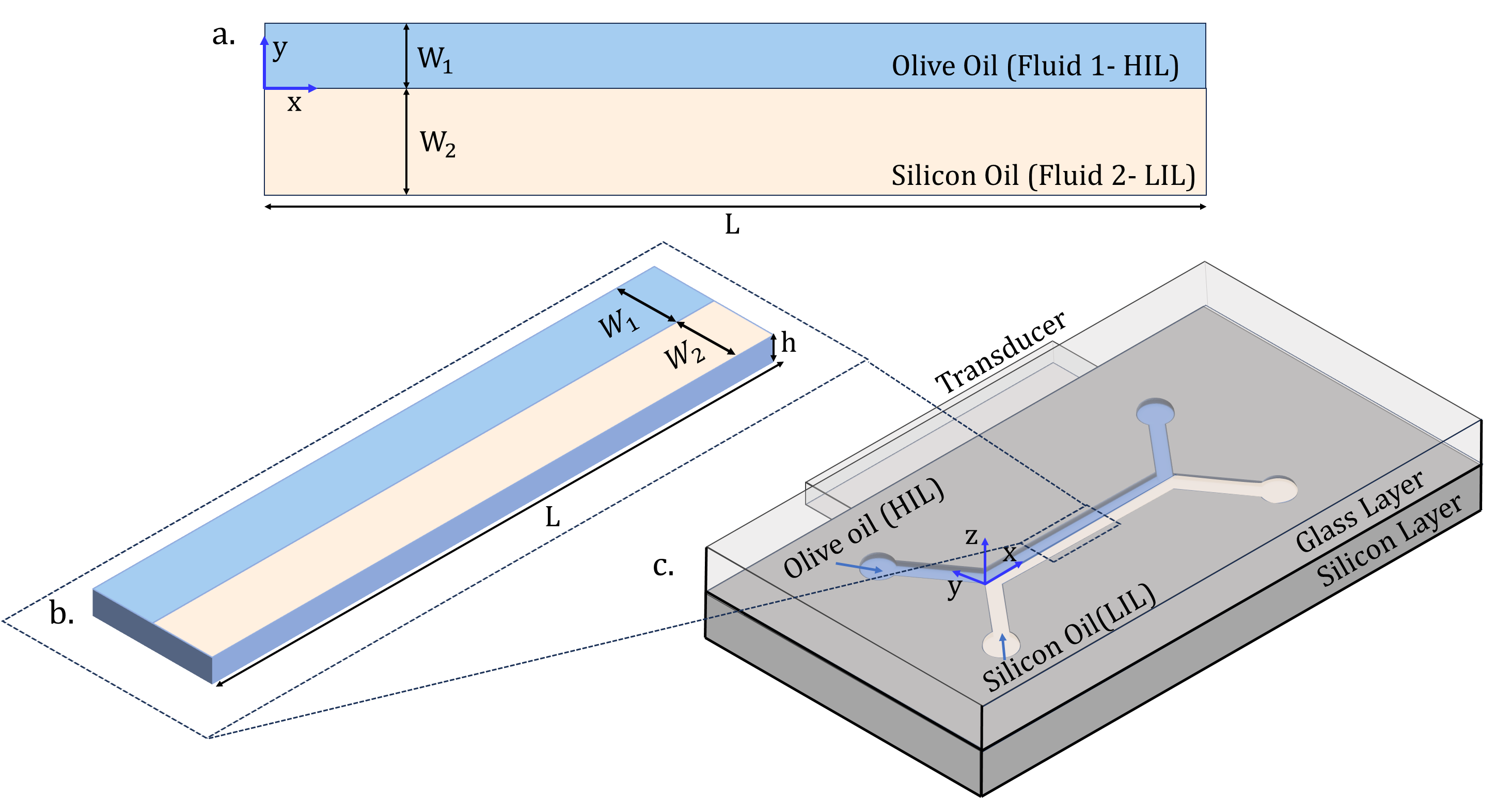}
\caption
{\justifying{ Geometries used in the numerical models. 
(a) Two-dimensional domain for dynamic phase-field simulations of coflowing immiscible liquids with widths $W_1$ (olive oil, HIL) and $W_2$ (silicone oil, LIL). 
(b) Reduced three-dimensional domain used to determine the position of the pressure nodal plane. 
(c) Full three-dimensional microfluidic device, including silicon and glass layers, used for chip acoustic simulations. 
The coordinate system is defined such that $x$, $y$, and $z$ denote the streamwise, transverse, and vertical directions, respectively.}}

\label{FigA1}
\end{figure}

 The geometries employed in the three numerical models used in this study is illustrated in Figure~\ref{FigA1}. Panel~(a) shows the two-dimensional domain used for the transient phase-field simulations, where two immiscible liquids flow in a coflow configuration with widths $W_1$ (Fluid~1, olive oil -- high acoustic impedance liquid, HIL) and $W_2$ (Fluid~2, silicone oil -- low acoustic impedance liquid, LIL). This model captures the interfacial dynamics and is described in \S~3.2. Panel~(b) presents the corresponding reduced three-dimensional domain used to determine the position of the pressure nodal plane, as discussed in \S~3.1. This domain retains the essential cross-sectional features of the channel while significantly reducing computational cost. Panel~(c) shows the full three-dimensional microfluidic device, including the silicon and glass layers, used for the complete chip-level acoustic simulations described in \S~3.3. The coordinate system is defined such that $x$, $y$, and $z$ denote the streamwise, transverse, and vertical directions, respectively.

\section{Regime map in the absence of acoustic actuation}\label{appB}

\begin{figure}
\centering
\includegraphics[clip,width=0.6\textwidth]{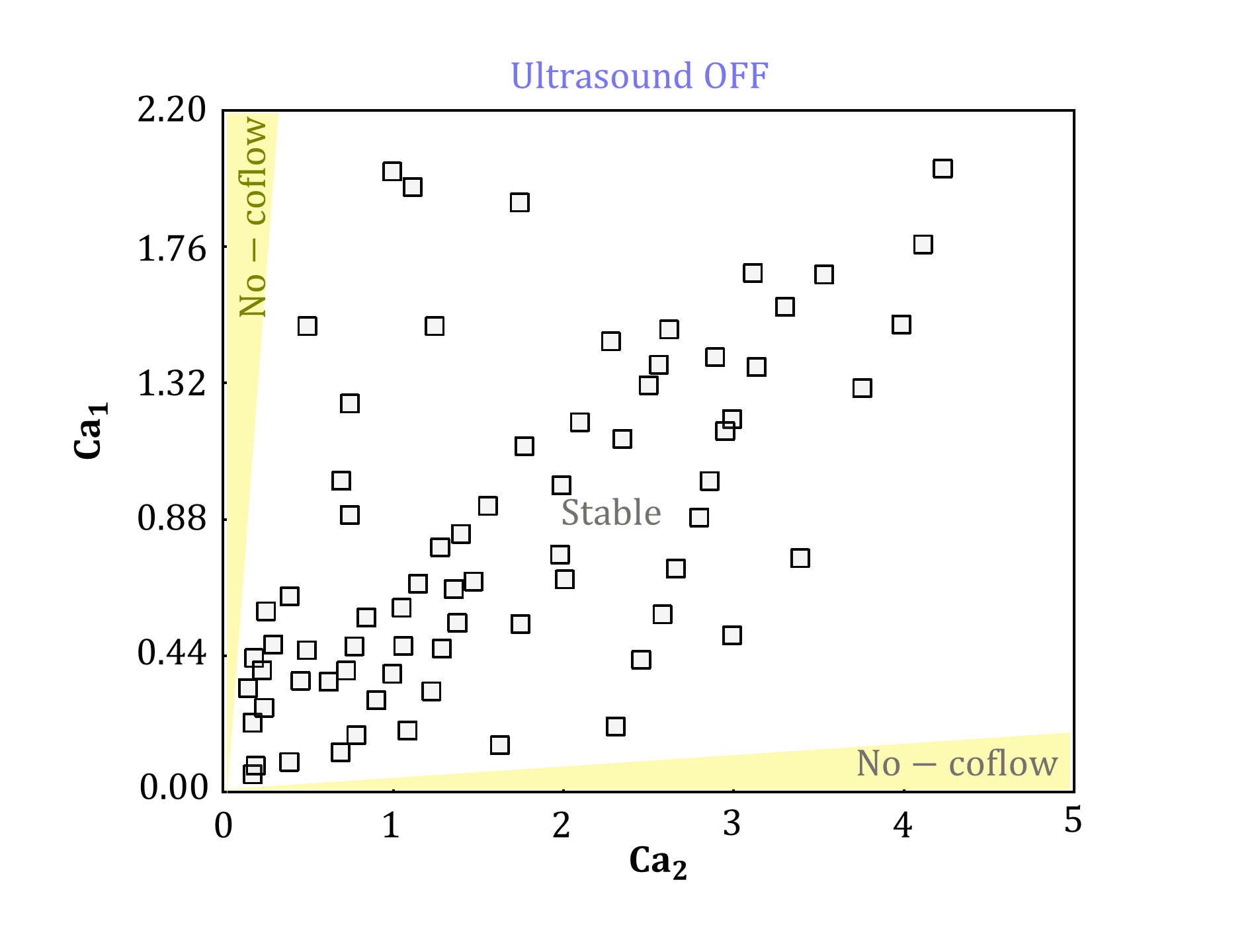}
\caption
{\justifying{Regime map in the absence of acoustic actuation ($V_{\mathrm{pp}}=0$), shown in terms of the capillary numbers $Ca_1$ and $Ca_2$. The system predominantly exhibits a stable coflow configuration over the explored parameter space, with no interfacial undulation. The yellow shaded regions indicate conditions where a parallel coflow cannot be established due to hydrodynamic imbalance.
}}

\label{FigA2}
\end{figure}

The regime map obtained in the absence of acoustic forcing is shown in Figure~\ref{FigA2}. The flow behaviour is characterized in terms of the capillary numbers of the two fluids, $Ca_1$ and $Ca_2$. In this case, the system predominantly exhibits a stable coflow configuration over a wide range of operating conditions, with no spontaneous deformation of the interface. The regions marked as “no coflow” correspond to conditions where a parallel two-stream configuration cannot be established due to strong hydrodynamic stress imbalance. These results establish the hydrodynamic baseline state of the system, highlighting that the observed instabilities reported in this work originates solely due to the presence of acoustic forcing that gets tuned via the hydrodynamics.

\section{Base coflow characteristics in the absence of acoustic forcing}\label{appC}

In the absence of acoustic excitation, the coflow remains stable and is taken as the base coflow. The velocity distribution in the two phases follows the analytical formulation of \cite{Hazra2024}, where the streamwise velocity in each fluid is described by a quadratic profile in the transverse coordinate $y$. The full expressions for the velocity field and the corresponding coefficients are provided in supplementary material S2.

\begin{figure}
\centering
\includegraphics[clip,width=0.8\textwidth]{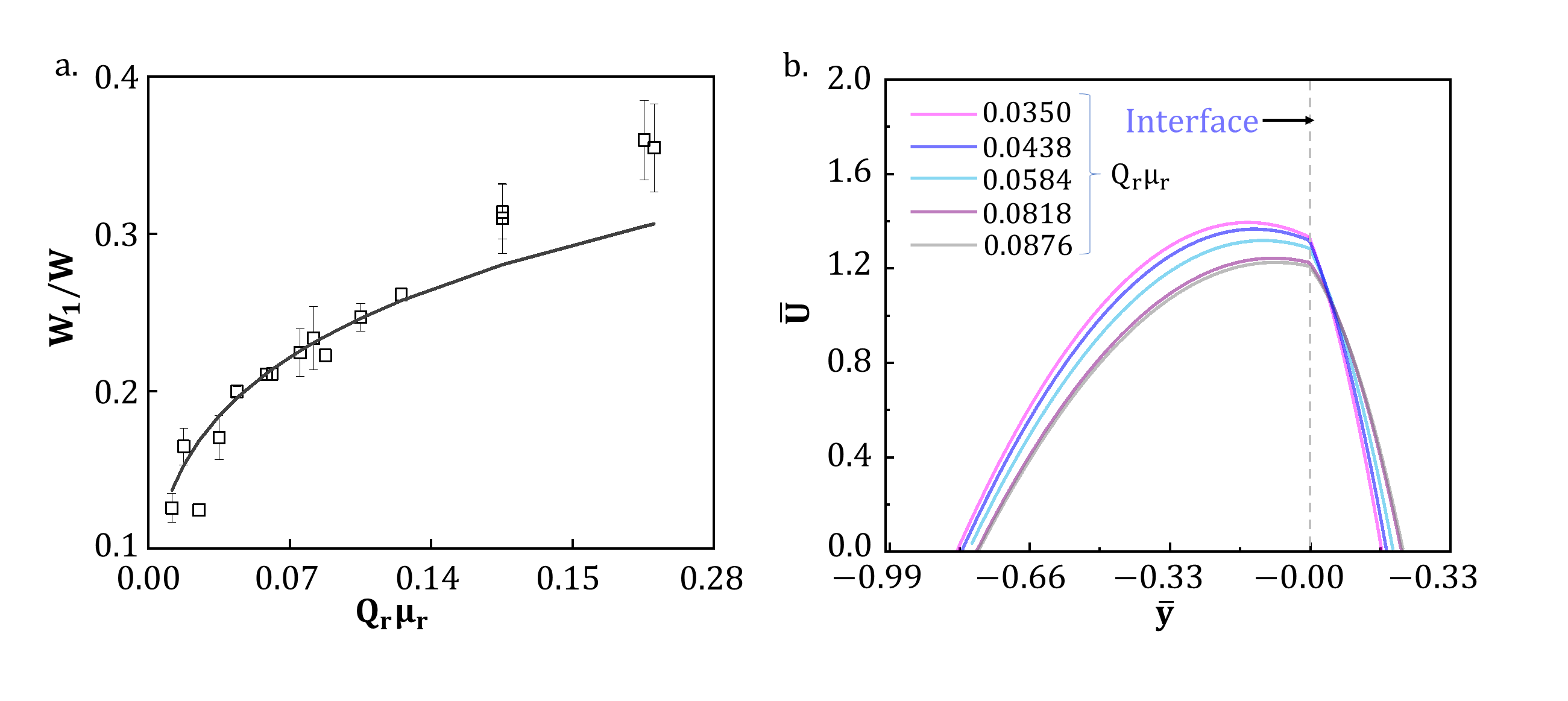}
\caption
{\justifying{
Base flow characteristics in the absence of acoustic forcing. 
(a) Variation of the normalized width $W_1/W$ with $Q_r\mu_r$. 
(b) Nondimensional velocity profiles $\bar{U}=U/U_0$ across the channel width for different $Q_r\mu_r$, plotted against $\bar{y}=y/W$; $\bar{y}=0$ (dashed line) denotes the interface location.
}}

\label{FigA3}
\end{figure}

The distribution of a pair of fluids across the channel is characterized by the width ratio $n = W_1/W_2$, which depends on the flow-rate ratio $Q_r = Q_1/Q_2$ and the viscosity ratio $\mu_r = \mu_1/\mu_2$. The normalized width of the high-impedance liquid (HIL) is given by $W_1/W = 1-(1+0.3(Q_r\mu_r)^{-1/3})^{-2}$. The variation of $W_1/W$ as a function of the combined parameter $Q_r\mu_r$ is shown in Figure~\ref{FigA3}(a), which demonstrates a monotonic increase with increasing hydrodynamic dominance of Liquid~1. The corresponding velocity profiles for different $Q_r\mu_r$ values is shown in Figure~\ref{FigA3}(b). The velocity is nondimensionalized as $\bar{U} = U/U_0$, where $U_0 = Q_0/A$ is the average velocity based on the total flow rate $Q_0 = Q_1 + Q_2$ and the cross-sectional area $A = W h$. The transverse coordinate is normalized as $\bar{y} = y/W$, with $\bar{y}=0$ (indicated by the vertical dashed line) representing the interface location. 

\section{Experimental results for weakly wall adhering lluid}\label{appd}

\begin{figure}
\centering
\includegraphics[clip,width=0.8\textwidth]{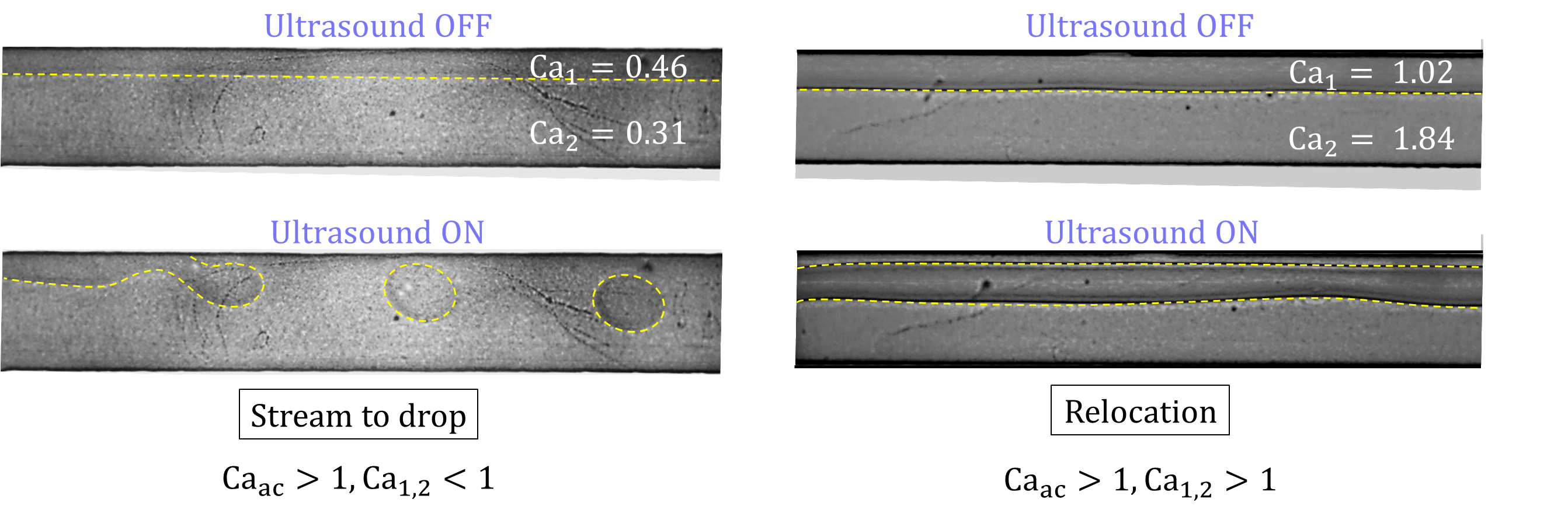}
\caption
{\justifying{
Flow behaviour for weakly wall-adhering liquids ($\theta_s \sim 90^\circ$) using an aqueous glycerol (80\%)--mineral oil pair. In the absence of acoustic forcing, a stable coflow is observed. Under acoustic excitation, the system exhibits direct stream-to-droplet breakup for $Ca_{1,2}<1$ and $Ca_{\mathrm{ac}}>1$, and interface relocation toward the pressure nodal plane for $Ca_{1,2}>1$ and $Ca_{\mathrm{ac}}>1$. No splitting is observed.}}
\label{FigA4}
\end{figure}

The flow behaviour for weakly wall-adhering liquids ($\theta_s \sim 90^\circ$), discussed in \S~4.2, is shown in Figure~\ref{FigA4}. The experiments are performed using an aqueous glycerol (80\%)--mineral oil pair, where the aqueous glycerol acts as the high-impedance liquid (HIL) and mineral oil as the low-impedance liquid (LIL). The aqueous glycerol phase has density $\rho_1 = 1207~\mathrm{kg\,m^{-3}}$, viscosity $\mu_1 = 47~\mathrm{mPa\cdot s}$, speed of sound $c = 1884 ~\mathrm{m\,s^{-1}}$ and inetrfacial tension $\gamma \approx 3\pm 0.15~\mathrm{mN\,m^{-1}}$.

In the absence of acoustic forcing, the system remains in a stable coflow configuration with a nearly flat interface. Upon application of acoustic forcing, the response depends on the capillary numbers. For $Ca_{1,2}<1$ and $Ca_{\mathrm{ac}}>1$, the interface undergoes direct stream-to-droplet breakup, whereas for $Ca_{1,2}>1$ and $Ca_{\mathrm{ac}}>1$, the interface relocates toward the pressure nodal plane without breakup. In contrast to a strongly wall-adhering coflow systems, no splitting is observed in a weakly wall-adhering coflow system.

\section{Schematic of interfacial protrusion and force balance}\label{appe}

\begin{figure}
\centering
\includegraphics[clip,width=0.6\textwidth]{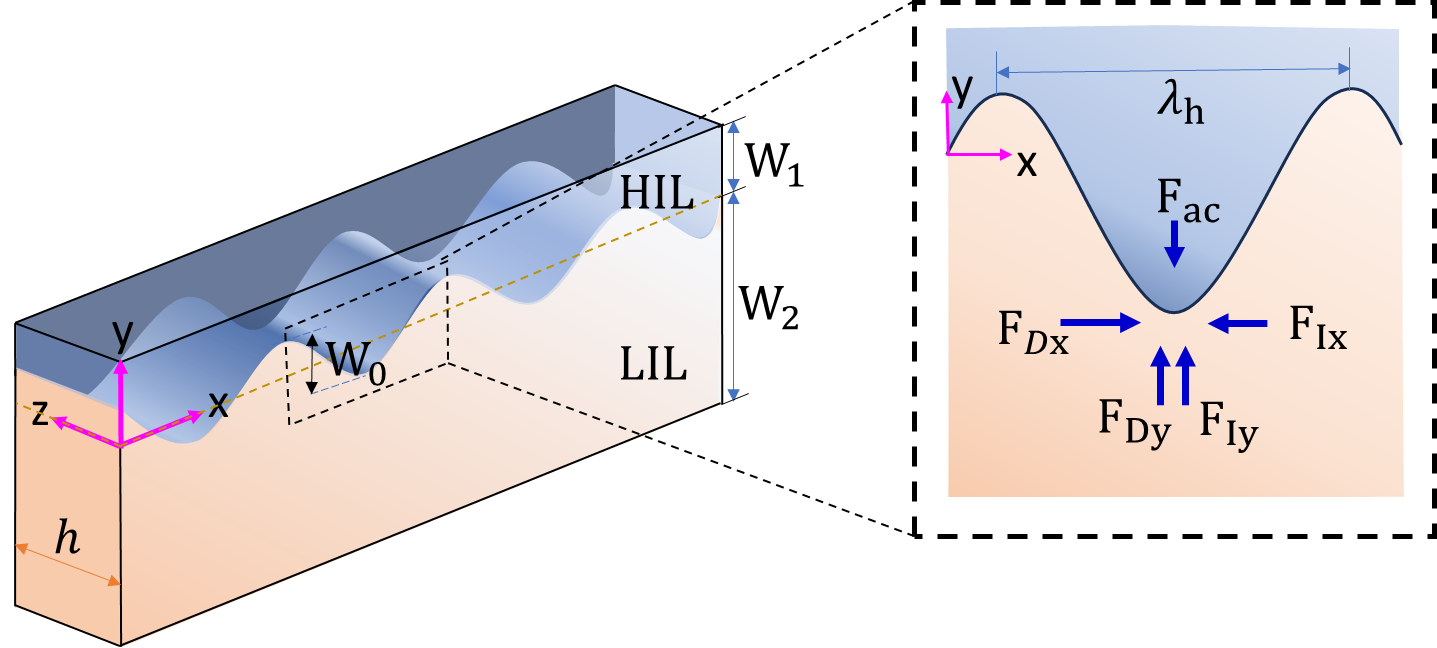}
\caption
{\justifying{
Schematic of the interfacial protrusion used in the scaling analysis of \S~4.3.1. The interface between the high-impedance liquid (HIL) and low-impedance liquid (LIL) undergoes a finite-amplitude deformation under acoustic forcing, with a characteristic streamwise length scale of order $\lambda_H$. The inset shows the control volume surrounding a typical protrusion, indicating the relevant forces: acoustic radiation force $F_{\mathrm{ac}}$ acting normal to the interface, viscous drag forces $F_{Dx}$ and $F_{Dy}$, and interfacial restoring forces $F_{Ix}$ and $F_{Iy}$ arising from interfacial tension.}}
\label{FigA5}
\end{figure}

A schematic representation of the interfacial protrusion considered in the scaling analysis of \S~4.3.1 is shown in Figure~\ref{FigA5}. The interface between the high-impedance liquid (HIL) and low-impedance liquid (LIL) develops a finite-amplitude deformation under acoustic forcing, forming a protrusion of characteristic streamwise length scale comparable to the acoustic wavelength $\lambda_h$. The enlarged view highlights the control volume surrounding a typical protrusion, on which the relevant forces are defined. The acoustic radiation force $F_{\mathrm{ac}}$ acts normal to the interface, driving the deformation. In the streamwise direction, the drag force $F_{D_x}$ is balanced by the interfacial restoring force $F_{I_x}$, while in the transverse direction the combined action of $F_{\mathrm{ac}}$ and $F_{D_y}$ is opposed by the interfacial restoring force $F_{I_y}$. This schematic provides a visual basis for the force balance used to estimate the transition between waviness and splitting regimes.

\section{Breakup location under varying hydrodynamic conditions}\label{appf}

\begin{figure}
\centering
\includegraphics[clip,width=0.5\textwidth]{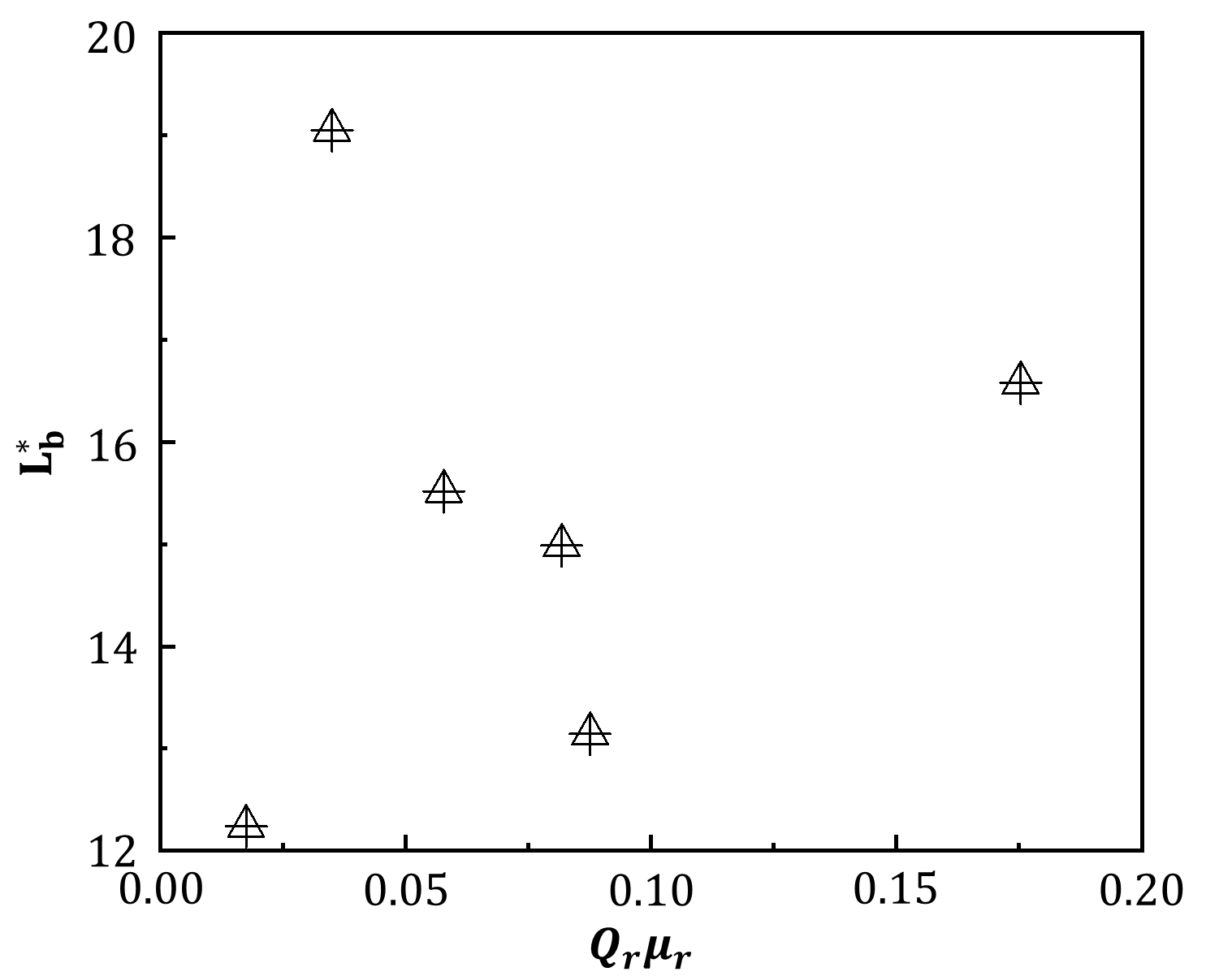}
\caption
{\justifying{
Variation of the normalized breakup length in the splitting regime with $Q_r\mu_r$ at fixed acoustic power. The data show significant scatter and no clear monotonic dependence on $Q_r\mu_r$.}}
\label{FigA6}
\end{figure}

The variation of the normalized breakup length in the splitting regime with the combined hydrodynamic parameter $Q_r \mu_r$ at fixed acoustic power is shown in figure~\ref{FigA6}. Here, $Q_r = Q_1/Q_2$ is the flow-rate ratio and $\mu_r = \mu_1/\mu_2$ is the viscosity ratio. The data correspond to different experimental conditions obtained by varying $Q_r$ and $\mu_r$ independently while maintaining a constant acoustic forcing. In contrast to the systematic dependence observed with acoustic power (see \S~4.4), the breakup location does not exhibit a clear monotonic variation with $Q_r \mu_r$. Instead, significant scatter is observed across the parameter range. This reflects the competing influence of flow-rate and viscosity contrast on the interfacial dynamics, indicating that the breakup location in the splitting regime cannot be uniquely characterized by the combined parameter $Q_r \mu_r$.

\section{Results of fully-coupled electromechanical simulation}\label{app:G}
To further examine the spatial structure of the acoustic forcing, fully-coupled electromechanical simulations are performed to resolve the first-order acoustic pressure field within the microchannel. The numerical framework, described in \S\ref{sec:3.1}, is used to capture the standing-wave field generated by the imposed actuation.

The simulations correspond to a representative experimental configuration consisting of a silicone oil (low-impedance liquid) and olive oil (high-impedance liquid) system, with initial widths $W_1 = 150~\mu\mathrm{m}$ and $W_2 = 220~\mu\mathrm{m}$. The device is actuated using a harmonic excitation with an applied voltage $V_{pp} = 5~\mathrm{V}$, which induces structural vibrations of the substrate and generates an acoustic field within the fluid domain. The resulting electromechanical response produces a maximum structural displacement amplitude of $V_s = 9.2~\mathrm{nm}$, which sets the strength of the resulting acoustic field.

The resulting first-order acoustic pressure field $p_1$ forms a standing-wave pattern inside the channel. The spatial distribution of $p_1$ is shown in Figure~\ref{FigA6}, where the white region corresponds to the pressure nodal plane and the red and blue regions represent the positive and negative extrema of the acoustic pressure amplitude. The pressure amplitude reaches approximately $\pm 1.6~\mathrm{MPa}$ in the present simulation.

To quantify the axial variation, the acoustic pressure is extracted along a line at the channel centre. The resulting profile exhibits a clear periodic variation along the streamwise direction, forming a wavy pattern with a well-defined wavelength. The extracted wavelength is $\lambda \approx 750~\mu\mathrm{m}$, which closely matches the acoustic wavelength imposed by the standing-wave field.

This axial periodicity of the acoustic pressure field is consistent with the wavelength of the interfacial deformation observed experimentally, confirming that the instability is governed by the spatial structure of the imposed acoustic field. Notably, this axial variation is not captured in the reduced-order model, which does not resolve the full streamwise structure of the acoustic field, and hence a fully-coupled electromechanical simulations is required.


\section{Theoretical estimation of the thin residual stream (TRS) thickness in the splitting regime}\label{appH}

We consider a co-flow configuration in which a high-impedance liquid (HIL) of velocity $U_1$ and width $W_1$ interacts with a low-impedance liquid (LIL) of velocity $U_2$ in a channel of height $h$ under acoustic actuation. Upon actuation, the HIL splits into droplets of characteristic diameter $d$, while the remaining portion forms a thin residual layer of thickness $t_a$ along the channel wall. The volumetric flow rates are $Q_1 = U_1 W_1 h$ and $Q_2 = U_2 W_2 h$. Under steady splitting conditions, the incoming HIL is partitioned between the residual layer and droplet formation such that $Q_1 = Q_t + Q_d$, where $Q_t = (U_1 + U_2) h t_a$ represents the transport of the residual layer. The droplet volume is estimated as $V_{\text{drop}} = (\pi d^2/4)h$. The droplet formation frequency is taken to scale with the relative convective velocity between the two streams, $f \sim |U_2 - U_1|/(2\lambda)$, reflecting the rate at which interfacial protrusions are advected and pinched off over a characteristic spacing set by the acoustic field. The droplet production rate is further influenced by the relative transport of the two phases. Since the LIL advects the interfacial protrusions downstream, the effective rate of HIL removal into droplets scales with the velocity ratio $U_2/U_1$, yielding
\begin{equation}\label{eq:E1}
Q_d = \left(\frac{U_2}{U_1}\right) V_{\text{drop}} f.
\end{equation}

Using $Q_1 = Q_t + Q_d$, the residual thickness is obtained as
\begin{equation}\label{eq:E2}
t_a = \frac{Q_1 - Q_d}{(U_1 + U_2)h}.
\end{equation}

Substituting $Q_1$ and $Q_d$ and normalizing with $W$ gives
\begin{equation}\label{eq:E3}
t_a^* = \frac{U_1 W_1}{W(U_1 + U_2)}
- \left(\frac{U_2}{U_1}\right)\frac{\pi d^2 |U_2 - U_1|}{8\lambda W (U_1 + U_2)}.
\end{equation}

Expressing $U_1 = Q_1/(W_1 h)$ and $U_2 = Q_2/(W_2 h)$ and introducing $Q_r = Q_1/Q_2$, we obtain
\begin{equation}\label{eq:E4}
t_a^* = \frac{Q_r W_1 W_2}{W(Q_r W_2 + W_1)}
- \frac{\pi d^2}{8\lambda W}
\left(\frac{Q_2 W_1}{Q_1 W_2}\right)
\left[\frac{|Q_2 W_1 - Q_1 W_2|}{Q_1 W_2 + Q_2 W_1}\right].
\end{equation}

Using $\beta = \left(1 + 0.3 (Q_r \mu_r)^{1/3}\right)^2$, with $W_2 = W/\beta$ and $W_1 = W(\beta - 1)/\beta$, the expression reduces to
\begin{equation}\label{eq:E5}
t_a^* = \alpha \left(
\frac{Q_r}{\beta} - 
\frac{\pi d^2 |\beta - 1 - Q_r|}{8\lambda W Q_r}
\right),
\end{equation}
where $\alpha = (\beta - 1)/(Q_r + \beta - 1)$.

This expression captures the reduction in residual-layer thickness due to droplet formation and is consistent with experimental observations. 

\counterwithin*{equation}{section}
\renewcommand\theequation{\thesection\arabic{equation}}

\bibliographystyle{jfm}
\bibliography{jfm}

\clearpage
\section*{Supplementary material}
\addcontentsline{toc}{section}{Supplementary material}

\setcounter{section}{0}
\setcounter{figure}{0}
\setcounter{table}{0}
\setcounter{equation}{0}

\renewcommand{\thefigure}{S\arabic{figure}}
\renewcommand{\thetable}{S\arabic{table}}
\renewcommand{\theequation}{E\arabic{equation}}

\newcommand{\suppsection}[1]{%
  \refstepcounter{section}%
  \section*{S\arabic{section}. #1}%
  \addcontentsline{toc}{section}{S\arabic{section}. #1}
}

\suppsection{Resonance characteristics, pressure nodal plane, and interfacial dynamics}\label{S1}

\begin{figure}
\centering
\includegraphics[clip,width=1\textwidth]{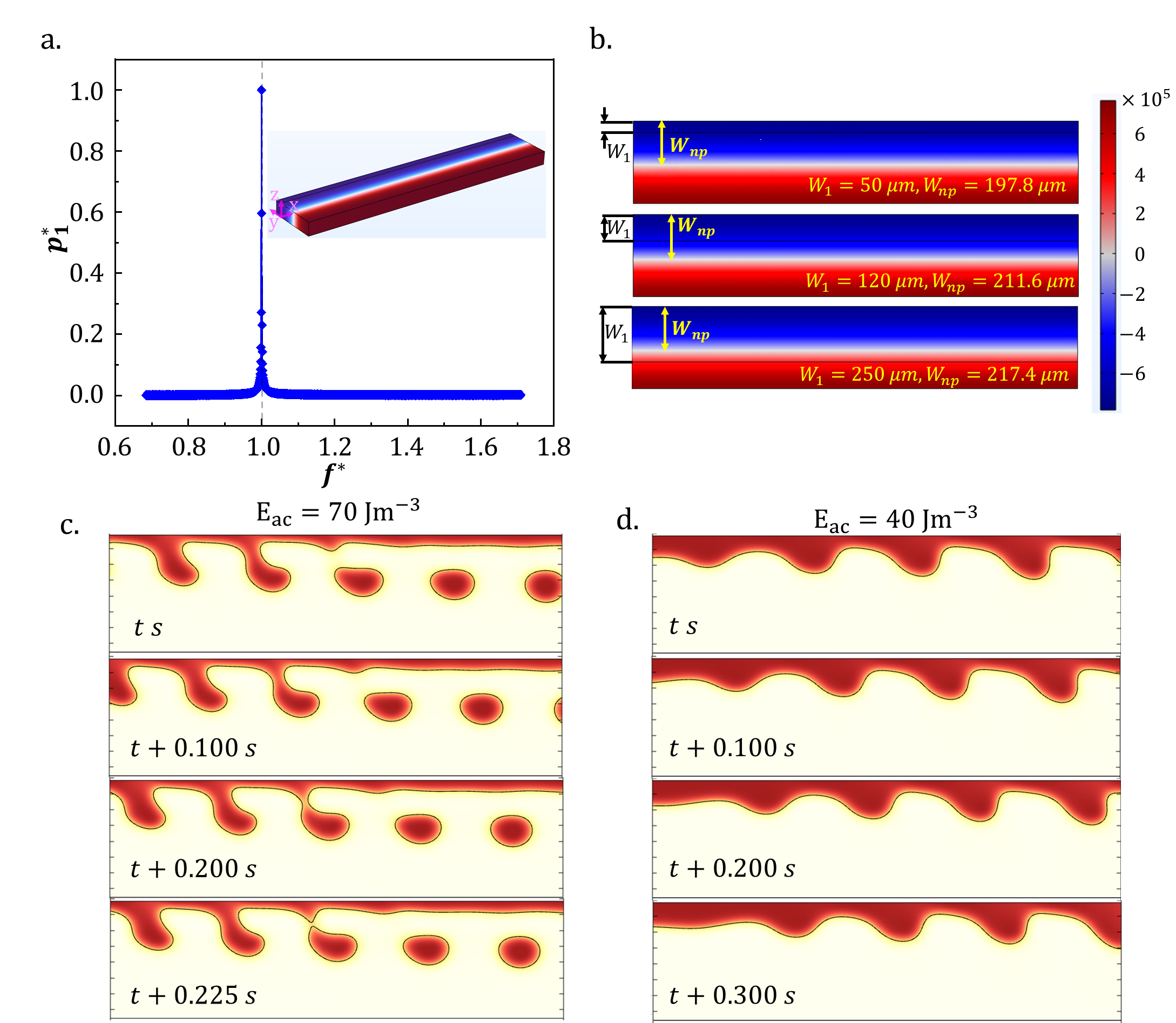}
\caption
{
\justifying{Numerical results illustrating resonance characteristics, pressure nodal plane location, and interfacial dynamics. All simulations correspond to a silicone oil (350 cSt)–olive oil coflow system. 
(a) Variation of the non-dimensionalised first-order acoustic pressure amplitude $p_1^*$ as a function of the normalised frequency $f^* = f/f_{\mathrm{res}}$, showing a sharp resonance peak at $f^* = 1$. 
(b) Spatial distribution of the acoustic pressure field showing the position of the pressure nodal plane ($W_{np}$) for different inlet widths $W_1$ (and corresponding $W_2$). The nodal plane ($p_1 = 0$) shifts with the geometric configuration of the coflow. 
(c,d) Time evolution of the interface for identical hydrodynamic conditions ($Ca_1 = 0.857$ and $Ca_2 = 1.6$) under different acoustic energy densities. (c) High acoustic forcing ($E_{ac} = 70~\mathrm{J/m^3}$) leading to the splitting regime, and (d) lower acoustic forcing ($E_{ac} = 40~\mathrm{J/m^3}$) resulting in the waviness regime.}}
\label{Sf1}
\end{figure}

To elucidate the role of acoustic forcing in the present system, we analyse the resonance characteristics, the spatial structure of the acoustic field, and the resulting interfacial dynamics using the reduced-order three-dimensional numerical model described in \S\,3 of the main text, together with full-device simulations. All results in Figure~\ref{Sf1} correspond to the silicone oil (350~mPa\,s)–olive oil system.

Figure~\ref{Sf1}(a) shows the variation of the non-dimensionalised first-order acoustic pressure amplitude, $p_1^* = p_1/p_{1,\max}$, as a function of the normalised frequency $f^* = f/f_{\mathrm{res}}$, where $f_{\mathrm{res}}$ is the resonance frequency obtained from the reduced-order eigenfrequency analysis. A sharp peak at $f^* = 1$ indicates resonance, correspondingly offers the maximum acoustic energy density within the microchannel. The resonance frequency predicted by the reduced-order model deviates from the experimentally observed value, owing to the simplified representation of the device, which does not fully account for electromechanical coupling and energy damping mechanisms. In contrast, full-device simulations that incorporate the complete geometry yield a resonant frequency response in closer agreement with experiments. Despite this discrepancy in frequency prediction, the reduced-order model accurately captures the spatial structure of the acoustic field, in particular the location of the pressure nodal plane defined by $p_1 = 0$. Figure~\ref{Sf1}(b) shows the position of the nodal plane ($W_{np}$) for different inlet widths $W_1$ (with corresponding $W_2$), demonstrating that its location varies systematically with the geometric configuration of the coflow. The liquid--liquid interface is established by hydrodynamic conditions, whereas the acoustic field imposes a spatially periodic forcing with a well-defined nodal structure. The relative position of the interface with respect to the nodal plane determines the symmetry of the acoustic forcing across the interface. When the interface is located close to the nodal plane, the acoustic forcing is nearly symmetric and results in weak deformation. As the gap between the two increases, a net transverse acoustic forcing develops, leading to sustained interfacial deformation.

The resulting interfacial dynamics, obtained from time-dependent simulations (\S\,3.3 in the main text), are shown in Figures~\ref{Sf1}(c,d) for identical hydrodynamic conditions ($Ca_1 = 0.857$ and $Ca_2 = 1.6$) but different acoustic energy densities. At higher acoustic energy density ($E_{ac} = 70~\mathrm{J/m^3}$), strong deformation leads to the splitting regime (Figure~\ref{Sf1}(c)), whereas at lower acoustic energy density ($E_{ac} = 40~\mathrm{J/m^3}$), the interface exhibits sustained waviness without breakup (Figure~\ref{Sf1}(d)).

The temporal evolution shows that the interface is initially stable under purely hydrodynamic conditions. Upon activation of the acoustic field, time-periodic acoustic radiation forces generate transverse interfacial perturbations. The growth of these perturbations is governed by a competition between acoustic forcing and interfacial tension, while their downstream evolution is controlled by the interplay between viscous drag and interfacial tension force. When the acoustic forcing is sufficiently strong to overcome interfacial resistance ($Ca_{ac} > 1$), the perturbations grow in amplitude and are advected downstream due to visscous drag, leading to progressive elongation of the interface. If the streamwise viscous drag is also sufficiently large, the elongated protrusions undergo necking and eventually pinch off to form droplets, resulting in the splitting regime. This process involves a sequence of transverse deformation, downstream advection, neck thinning, and breakup, followed by the formation of a quasi-periodic droplet train along with thin residual stream. In contrast, when acoustic forcing is sufficient to deform the interface but the streamwise drag remains moderate, the perturbations do not undergo pinch-off. Instead, a sustained wavy interface is observed, corresponding to the waviness regime. 

These results demonstrate that the observed regimes arise from the coupled interplay between acoustic radiation forces, interfacial tension, and viscous drag asymmetry, consistent with the mechanism described in the main text. The reduced-order model captures the essential spatial structure of the acoustic field, while the full-device simulations provide improved quantitative agreement with experimental observations.

\suppsection{ Base flow formulation and width ratio prediction}\label{S2}

In the absence of acoustic excitation, the coflow of the two immiscible liquids remains stable and serves as the base flow for the present analysis. The velocity profile of the coflow is described using the formulation of ~\cite{Yih_1967, Hazra2024}, where the streamwise velocity profiles of the two phases are expressed as quadratic functions of the transverse coordinate $y$. The velocity profiles in the two phases are given by
\begin{equation}\label{eq:S4_1}
U_1 = A_1 y^2 + a_1 y + b,
\end{equation}
\begin{equation}\label{eq:S4_2}
U_2 = A_2 y^2 + a_2 y + b,
\end{equation}
where $U_1$ and $U_2$ denote the velocities in phases 1 and 2, respectively. The coefficients $A_1$, $A_2$, $a_1$, $a_2$, and $b$ are determined from the continuity of velocity and shear stress at the interface, along with the imposed pressure gradient. These coefficients are given by \citep{Yih_1967,Hazra2024}
\[
A_1 = \frac{A_2}{\mu_r}, \quad A_2 = -\frac{R_2 K}{2},
\]
\[
a_1 = \frac{K R_2 (\mu_r - n^2)}{2 \mu_r (\mu_r + n)}, \quad a_2 = \mu_r a_1,
\]
\[
b = \frac{K R_2 (n + n^2)}{2 (\mu_r + n)},
\]
where $\mu_r = \mu_1/\mu_2$ is the viscosity ratio, $Q_r = Q_1/Q_2$ is the flow rate ratio, and $n = W_1/W_2$ is the width ratio of the two streams. The Reynolds numbers are defined as $R_1 = U_0 \rho_1 W_2/\mu_1$ and $R_2 = U_0 \rho_2 W_2/\mu_2$, where $U_0$ is the average velocity in the channel.

The pressure gradient parameter $K$ is given by~\cite{Yih_1967, Hazra2024}
\begin{equation}\label{eq:S4_3}
K = \frac{(1 + n)(Q_2 / Q_t)}{R_2 \left[ -\frac{1}{6} + \frac{\mu_r + 2n + n^2}{4\mu_r + 4n} \right]},
\end{equation}
where $Q_t$ is the total flow rate in the channel. To determine the width ratio $n$, the average velocity in phase 1 is obtained from equation~(\ref{eq:S4_1}) and multiplied by the cross-sectional area $(W_1 h)$ to compute the corresponding flow rate. By equating this with the imposed flow rate $Q_1$, the governing equation for $n$ reduces to a fourth-order algebraic equation~\cite{Yih_1967,Hazra2024}:
\begin{equation}\label{eq:S4_4}
n^{4} + 4\mu_r n^{3} - 3\mu_r n^{2}(Q_r - 1) - 4\mu_r Q_r n - \mu_r^{2} Q_r = 0.
\end{equation}

The solution of this equation yields the width ratio $n$, which can be expressed as \citep{Yih_1967,Hazra2024}
\begin{equation}\label{eq:S4_5}
n = \frac{1}{2} \left( -2\mu_r + \sqrt{G_1} + \sqrt{G_2} \right),
\end{equation}
where the intermediate quantities $G_1$, $G_2$, and $G_3$ are given by
\[
G_1 = \mu_r + 4\mu_r^2 + 3\mu_r(-1 + Q_r) - \mu_r Q_r + \frac{\mu_r^2(1 + Q_r)^2}{G_3} + G_3,
\]
\[
G_2 = -\mu_r + 8\mu_r^2 + 3\mu_r(-1 + Q_r) + \mu_r Q_r - \frac{\mu_r^2(1 + Q_r)^2}{G_3} - G_3 - \frac{4\mu_r(4\mu_r^2 + 3\mu_r(-1 + Q_r) - 2Q_r)}{\sqrt{G_1}},
\]
and
\[
G_3 = \left[ -8\mu_r^4 Q_r - \mu_r^3 Q_r^3 - 9\mu_r^3 Q_r^2 + 9\mu_r^3 Q_r + \mu_r^3 + 8\mu_r^2 Q_r^2 \right.
\]
\[
\left. - \left\{ \mu_r^4 \left( -\mu_r^2(1 + Q_r)^6 + \left(8\mu_r^2 Q_r - 8 Q_r^2 + \mu_r(-1 - 9Q_r + 9Q_r^2 + Q_r^3)\right)^2 \right) \right\}^{1/2} \right]^{1/3}.
\]
Once $n$ is obtained, the normalized width of phase 1 can be expressed as, $W_1/W = n / (1 + n)$. 

Plotting a continuous curve for width ratio in terms of viscosity and flow rate ratio is mathematically challenging. Nevertheless, using the above theoretical model, we compute discrete values of width ratio for specific values of viscosity and flow-rate ratios, over the experimental range. The discrete data obtained from the theoretical formulations is then utilized to develop a simplified empirical correlation given as, $W_1/W \approx 1 - \left(1 + 0.3 (Q_r \mu_r)^{1/3}\right)^{-2}$. A comparison between this empirical correlation curve and experimental measurements is presented in Appendix~C of the main text, showing good agreement over the explored parameter range.
\suppsection{Contact angle measurements and hysteresis}\label{S3}

\begin{figure}
\centering
\includegraphics[clip,width=0.7\textwidth]{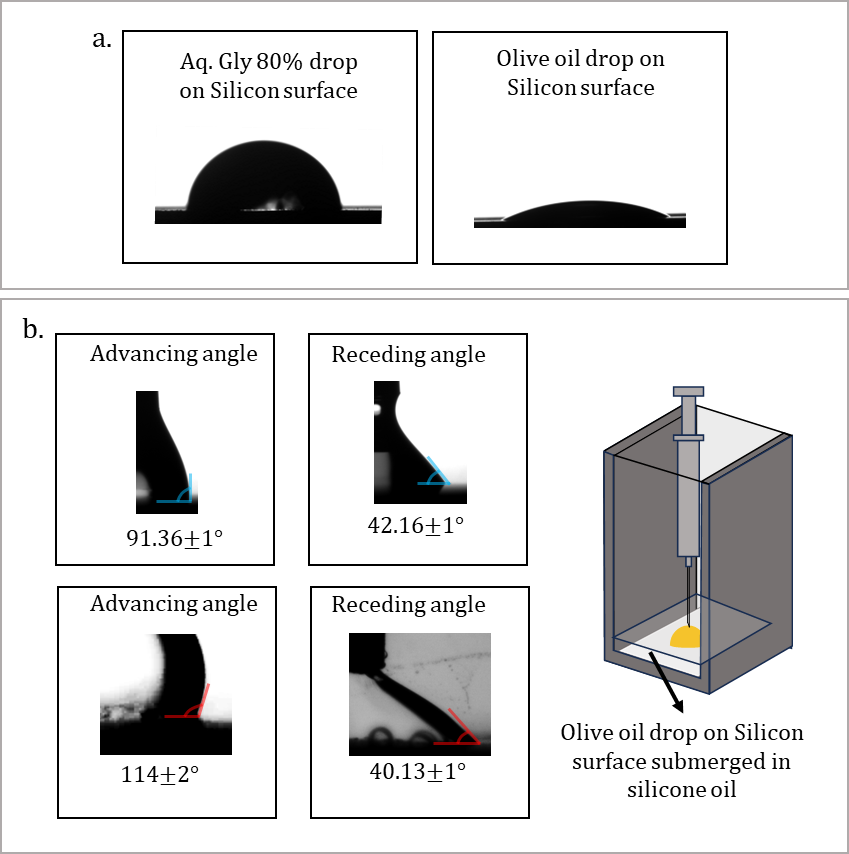}
\caption{
\justifying{Contact angle measurements for different fluid–solid combinations. 
(a) Static contact angle measured on a dry silicon surface in air. 
(b) Advancing and receding angles measured on a silicon surface submerged in silicone oil, depicting the experimental configuration used to replicate the microchannel environment. The advancing angle is obtained during dispense (pushing), while the receding angle is obtained during withdrawal (pulling). A schematic of the measurement setup is also shown.}
}
\label{Sf2}
\end{figure}

To quantify the role of wall adhesion and contact-line behaviour in the coflow system, advancing and receding angles were measured for the relevant fluid–solid combinations. These measurements provide a direct estimate of contact angle hysteresis, which characterises contact-line pinning and interfacial mobility along the channel walls. Figure~\ref{Sf2}(a) shows the static contact angle measured on a silicon surface in air for different liquid combinations. These values provide a baseline characterization of wettability and are reported in Table~1 of the main text. To observe the interfacial conditions inside the microchannel, advancing and receding angles were measured on a silicon surface submerged in silicone oil, as shown in Figure~\ref{Sf2}(b). A square silicon wafer was placed at the base of a cuvette filled with silicone oil (low-impedance liquid). The high-impedance liquid (olive oil or aqueous glycerol solution) was introduced using a syringe positioned close to the silicon surface, forming a three-phase (solid–liquid–liquid) contact line. The advancing angle was obtained by continuously dispensing high-impedance liquid, causing the contact line to move forward along the surface. The receding angle was measured by withdrawing liquid, leading to retreat of the contact line. The measurements were recorded using a goniometer-based imaging system.
For the silicon–olive oil system which exhibits splitting, the advancing angle was approximately $114^\circ$, while the receding angle was approximately $42^\circ$, resulting in a hysteresis of about $72^\circ$. In contrast, for the silicon–aqueous glycerol system which does not exhibit splitting, the advancing angle was approximately $91^\circ$ and the receding angle was approximately $40^\circ$, giving a hysteresis of about $51^\circ$. These results indicate that the silicon–olive oil system exhibits significantly larger hysteresis. Physically, larger hysteresis corresponds to stronger contact-line pinning and increased resistance to interfacial motion along the wall. In the strongly hysteretic silicon–olive oil system, acoustic forcing does not readily induce bulk lateral relocation of the stream. Instead, deformation becomes localized, promoting neck formation and eventual splitting. In contrast, the lower hysteresis observed for the silicon–aqueous glycerol system indicates weaker pinning, allowing the interface to relocate more readily under acoustic forcing without undergoing localized pinch-off.

These results demonstrate that contact angle hysteresis, rather than static wettability, plays a key role in determining the influence of wall interactions on the observed flow regimes.
\begin{figure}
\centering
\includegraphics[clip,width=1\textwidth]{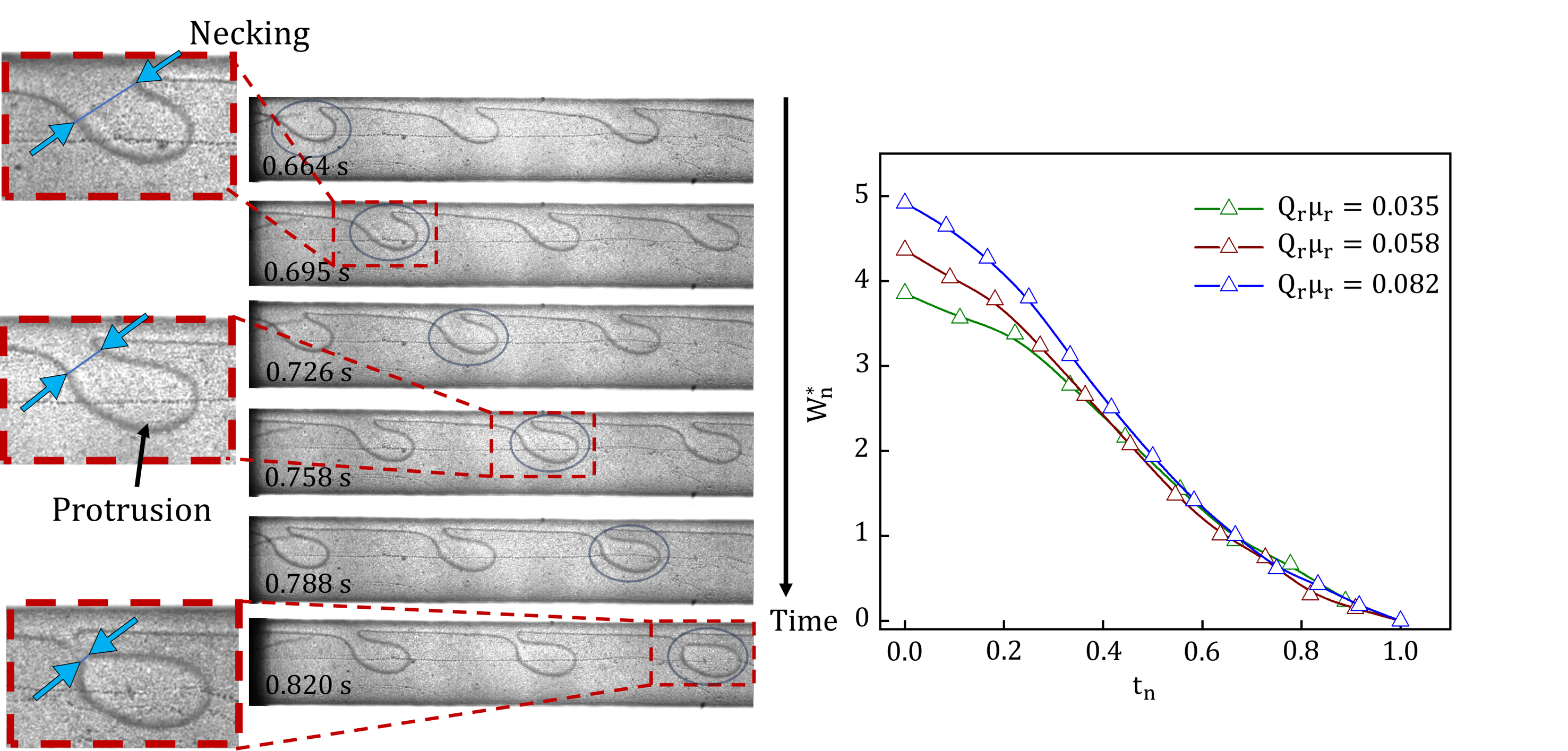}
\caption{\justifying{
Experimental observation of necking dynamics during stream splitting. Time-resolved snapshots showing the evolution of interfacial protrusions and the formation of a thinning neck prior to pinch-off. Insets highlight the neck region, and arrows indicate the measurement direction normal to the local interface. 
The plot shows the temporal evolution of the nondimensional neck width $W_n^*=W_n/W_1$ as a function of nondimensional time $t_n = t/t_b$ for different values of $Q_r \mu_r$, where $t_b$ denotes the breakup time. Higher $Q_r \mu_r$ corresponds to a wider initial neck that progressively thins to zero at pinch-off, leading to larger droplet sizes.}
}
\label{Sf3}
\end{figure}

\suppsection{Experimental observation of necking during splitting}\label{S4}

During the splitting regime, the acoustic field generates periodic interfacial undulations through time-periodic acoustic radiation forcing. These undulations form localized protrusions along the interface, whose spatial periodicity is set by the imposed acoustic field. As the deformation develops, the protrusions undergo streamwise elongation due to viscous drag asymmetry between the two liquid streams. This subsequently undergoes progressive thinning under the combined action of drag and interfacial tension, ultimately resulting in pinch-off and droplet formation accompanied by a thin residual stream (TRS). To quantify this process, we define the neck width $W_n$ as the minimum transverse thickness of the liquid filament, measured perpendicular to the local interface at the narrowest section immediately upstream of pinch-off. As illustrated in Figure~\ref{Sf3}, the measurement is taken along a direction normal to the interface contour, ensuring an accurate representation even when the filament is inclined relative to the channel axis. The neck width is nondimensionalized by the width of the high-impedance liquid (HIL), and is denoted as $W_n^*$. The temporal evolution of $W_n^*$, from the onset of neck formation to pinch-off, is shown in Figure~\ref{Sf3} for different values of the parameter $Q_r \mu_r$, which characterizes the relative strength of viscous stresses between the two streams. In all cases, the neck width decreases monotonically with time and approaches zero at the breakup point, consistent with the thinning dynamics leading to droplet formation.

Larger values of $Q_r \mu_r$ correspond to stronger viscous drag asymmetry, which enhances streamwise elongation of the protrusions and results in a wider neck at early times. Conversely, lower values of $Q_r \mu_r$ produce weaker elongation and thinner necks. These trends are consistent with the mechanism described in the main text, where viscous drag asymmetry governs the extent of elongation prior to pinch-off, while interfacial tension resists thinning. The time-resolved experimental images shown in Figure~\ref{Sf3} illustrate the sequence of events, including acoustically induced undulation, drag-driven elongation, neck thinning, and eventual pinch-off. Together, these observations provide direct experimental support for the proposed mechanism. A detailed analysis of the resulting droplet size and its dependence on flow conditions is presented in \S~4.8 of the main text.

\begin{figure}
\centering
\includegraphics[clip,width=0.8\textwidth]{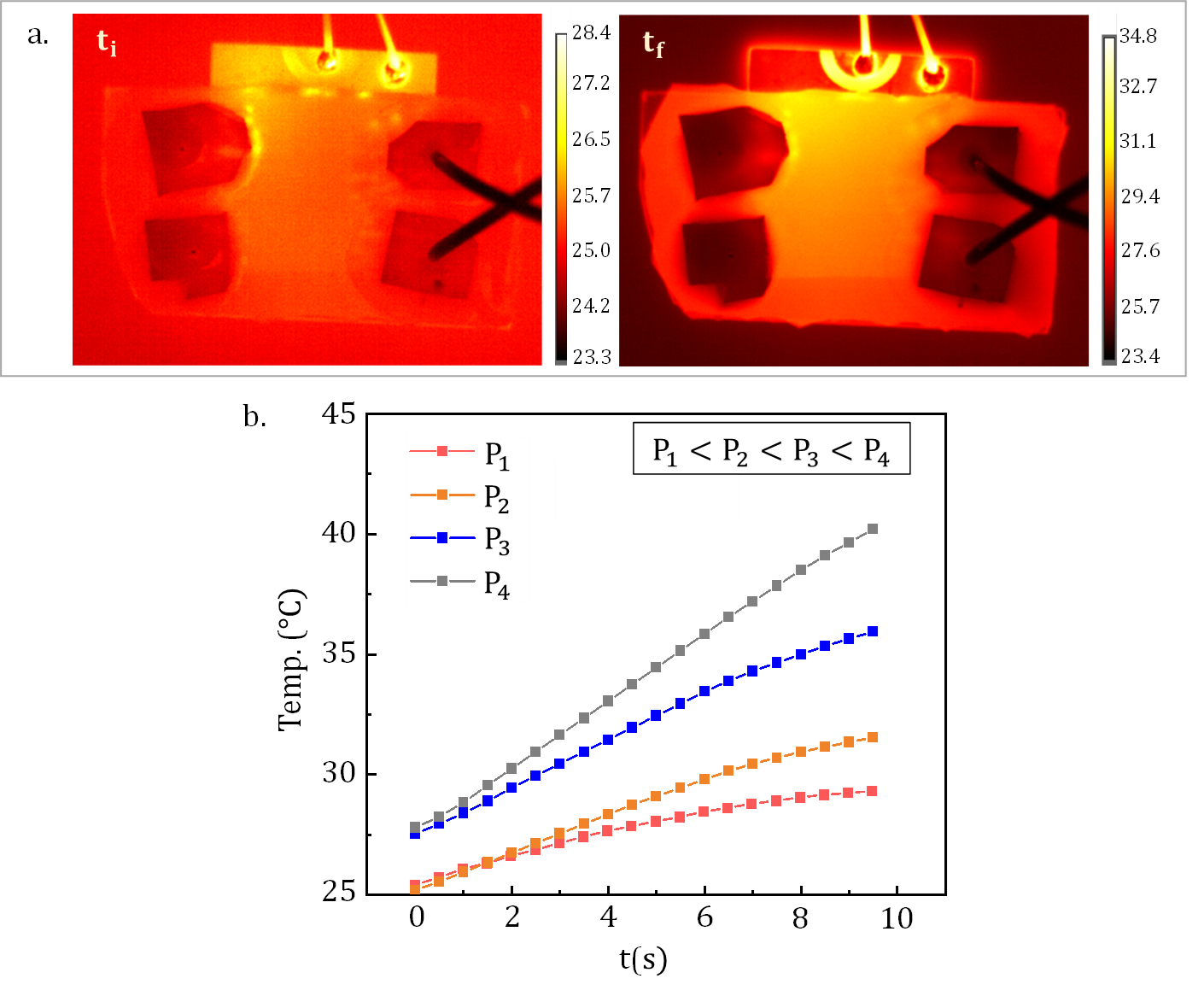}
\caption{
\justifying{(a) Infrared (IR) images of the device at an initial time $t_i$ and after ultrasound actuation at time $t$, showing a gradual and spatially uniform temperature rise. 
(b) Temporal evolution of temperature measured at four acoustic power levels ($P_1 < P_2 < P_3 < P_4$), demonstrating an approximately linear increase with time. The data are ensemble-averaged over three independent measurements, and the variation of temperature remains small ($<2^\circ$C) over the experimental timescale of 0.1-1.0 s.}
}
\label{Sf4}
\end{figure}
\suppsection{Assessment of thermal effects during acoustic actuation}\label{S5}

To assess the influence of thermal effects during acoustic actuation, we measured the transient temperature evolution of the device using an infrared (IR) camera. The objective was to determine whether heating could significantly alter fluid properties such as viscosity, density, or interfacial tension, and thereby influence the observed interfacial dynamics. In the present setup, the primary source of heating is the piezoelectric transducer due to electromechanical dissipation. The generated heat is conducted through the substrate and subsequently transferred to the fluid. As a result, the temperature rise within the fluid domain is indirect and governed by thermal diffusion from the solid substrate.

Figure~\ref{Sf4}(a) shows representative IR images of the device at the initial time $t_i$ and after a finite duration of actuation $t_f$, illustrating the spatial distribution of temperature. To quantify this behaviour, the temporal evolution of temperature was measured at four different acoustic power levels, denoted as $P_1 < P_2 < P_3 < P_4$, as shown in Figure~\ref{Sf4}(b). For each power level, three independent measurements were performed and ensemble-averaged, with the standard deviation used to estimate experimental variability. The results show that the temperature increases approximately linearly with time over the duration of measurement.

The measured heating rates were approximately $0.39~^\circ\mathrm{C\,s^{-1}}$ for moderate actuation ($P_1$=125~mW) and $0.64~^\circ\mathrm{C\,s^{-1}}$ for higher actuation ($P_2 = 220$~mW), with standard deviations below $0.3~^\circ\mathrm{C}$ over a $10~\mathrm{s}$ interval. These power levels correspond to the actuation conditions primarily used in the main experiments, and the reported statistics are representative of those cases. 

Over the characteristic experimental timescales (0.1--1.0~ms) relevant to interfacial deformation and breakup, the corresponding temperature rise remains small ($<2^\circ$C) even at higher acoustic power levels. This indicates that thermal effects evolve on a much longer timescale than the interfacial dynamics and therefore do not significantly influence the observed behaviour. The interfacial dynamics reported in the main text are thus governed primarily by acoustic forcing and hydrodynamic interactions, rather than by thermally induced variations.

\begin{figure}
\centering
\includegraphics[clip,width=0.8\textwidth]{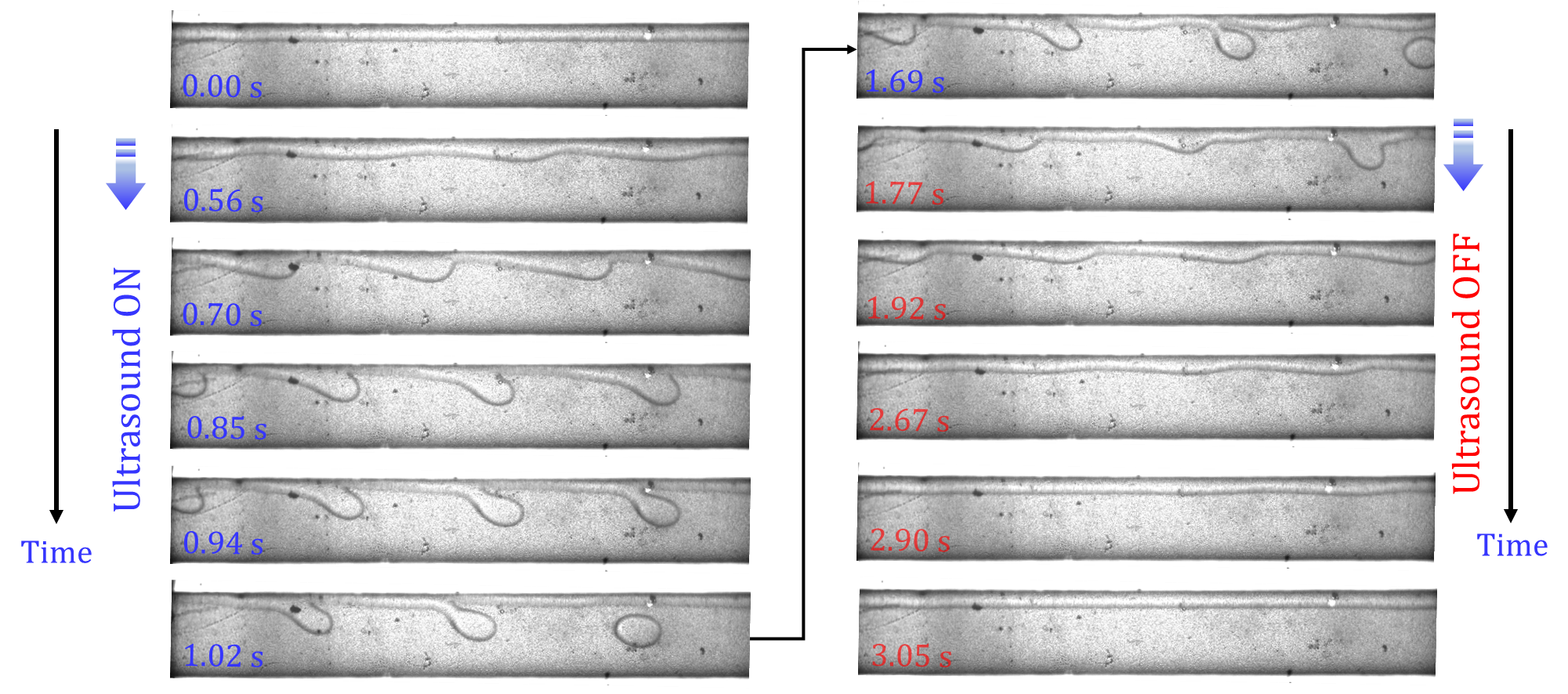}
\caption{
\justifying{Time-resolved experimental sequence demonstrating reversible transition between stable coflow and the splitting regimes. The acoustic field is switched on (left column), leading to the growth of interfacial perturbations and eventual breakup, and subsequently switched off (right column), resulting in relaxation back to a stable coflow.}
}
\label{Sf5}
\end{figure} 

\suppsection{Reversible transition between stable co-flow and splitting regime} \label{S6}

To demonstrate the reversible nature of the acoustically induced interfacial instability, we present a time-resolved sequence illustrating the transition between a stable coflow and the splitting regime, controlled solely by the application and removal of acoustic forcing. The experiment is performed at fixed flow conditions corresponding to a specific capillary number combination for which the coflow remains stable in the absence of acoustic excitation. As shown in Figure~\ref{Sf5}, the system initially exhibits a flat and undisturbed interface, characteristic of a stable parallel coflow of the two immiscible liquids. Upon activation of the acoustic field, an acoustic radiation force acts normal to the interface, initiating periodic interfacial perturbations. As the deformation develops, the protrusions undergo necking and eventual pinch-off, resulting in the breakup of the high-impedance liquid stream into discrete droplets accompanied by a thin residual film, consistent with the splitting regime.

When the acoustic field is switched off, the external forcing is removed and the interfacial deformation ceases. The interface rapidly relaxes back to its original flat configuration, restoring the stable co-flow. No residual deformation or hysteretic behaviour is observed in experimental observation.

This demonstrates that the transition between the stable and splitting regimes is sharp, controllable, and fully reversible. The results confirm that the instability is governed by externally imposed acoustic forcing, rather than by irreversible changes in the flow configuration or material properties.

\end{document}